\documentclass[aps,prb,reprint]{revtex4-2}
\usepackage{amsmath,amssymb} 
\usepackage{amsmath} 
\usepackage{amsfonts} 
\usepackage{bm} 
\usepackage[dvipdfmx]{graphicx} 
\usepackage{braket} 
\usepackage{physics}
\usepackage{color}
\usepackage{comment}
\usepackage{mathtools}

\begin{document}
\title{Spin amplitude wave due to dipole-quadrupole hybridization in pseudospin-1 pyrochlore magnets}

\author{Hiroki Nakai}
\email{nakai-hiroki3510@g.ecc.u-tokyo.ac.jp}
\author{Chisa Hotta}
\affiliation{Graduate School of Arts and Sciences, University of Tokyo, Meguro-ku, Tokyo 153-8902, Japan}

\date{\today}

\begin{abstract} 
We explore the quantum pseudospin-1 pyrochlore magnet featuring Fe$^{2+}$-based spinel oxides 
that addresses the formation of amplitude-modulated spin-density waves. 
We propose that the relatively small spin-orbit coupling and the small extra crystal field splitting 
in these materials create anisotropic exchange interactions and strong single-ion anisotropy, respectively, 
whose interplay becomes the source of quadrupolar moments selectively appearing on certain sublattices, 
leading to a spatially modulated hybrid of dipolar and quadrupolar moments. 
This mechanism represents the possibility of insulating magnets to form an exotic phase 
with coexisting liquid-solid properties. 
\end{abstract} 

\maketitle
\section{Introduction}\label{sec:intro}
One of the intriguing realizations of exotic material phases 
is the coexistence of liquid and solid components arising from the same degrees of freedom. 
This coexistence is exemplified by a supersolid phase, a coexisting long-range order of 
solid and superfluid component \cite{Andreev1969,Leggett1970} 
tested a long time ago in liquid He$^4$ \cite{Kim2004}, 
though it has remained experimentally elusive \cite{Kim2012} until finally  
the manipluation of Bose gas became possible in optical lattices \cite{Lonard2017,Li2017,Tanzi2019,Norcia2021}. 
In insulating quantum magnets, the supersolid phase is read off as a coexisting diagonal 
and off-diagonal long-range order for spin-1/2 degrees of freedom \cite{Matsuda1970, Liu1973}, 
and theoretically, a similar picture was studied in the context of geometrical frustration \cite{Wessel2005,Heidarian2005,Melko2005,Boninsegni2005,Hotta2006,Heidarian2010,Yamamoto2014,Yamamoto2015,Sellmann2015}. 
Still, providing experimental proofs is an ongoing challenge \cite{Xiang2024}. 
Meanwhile, ``magnetic moment fragmentation" in pyrochlore magnets was recently proposed as a realistic phenomenon, 
involving partial monopole crystallization within a disordered (classical liquid) spin ice medium 
\cite{Holdsworth2014, Petit2016, Paddison2016, Lefrancois2017, Dun2020}. 
These developments underscore the importance of understanding when and how such coexistence can be achieved. 
\par
Here we investigate the quantum pseudospin-1 pyrochlore magnet as a candidate for another type of coexisting phase: 
dipole and quadrupole taking the role of solid (magnetic) and liquid (nonmagnetic) components, respectively
\footnote{In the present case, the quadrupolar moment is induced by the on-site term, $D_z$ in Eq.(1), 
which works as a conjugate field to a quadrupole. 
In that respect, such a quadrupolar state is not a symmetry-breaking long-range order and it is commonly referred to as the quantum paramagnet. 
If we focus on the magnetic degrees of freedom, the only difference from the conventional paramagnet is that it is stabilized not by the thermal fluctuation but by the specific quantum fluctuation. 
In that context, it is more naturally referred to as ``liquid component".
}. 
Our model features spinel oxides such as $\rm GeFe_2O_4$ and $\gamma$-$\rm SiFe_2O_4$ \cite{Perversi2018}, 
which are reported to host amplitude-modulated spin-density waves (SDWs). 
In these compounds, $\rm Fe^{2+}$ ions form a pyrochlore lattice, where each 
$d^6$ electron configuration produces a spin $S=2$ with $L=2$, which, 
in the cubic crystal field reduces to the $^5T_{2g} (t_{2g}^4 e_g^2)$ multiplet with effective orbital moment $L_{\rm eff}=1$. 
Importantly, these materials do not exhibit any lattice distortion associated with the magnetic transition \cite{Perversi2018,Barton2014}, 
which is unusual in spinel compounds with orbital degeneracy. 
It is then supposed that the orbital fluctuation produces the spatially nonuniform 
exchange interactions from ferromagnetic to antiferromagnetic, which is responsible for the observed SDW \cite{Perversi2018}. 
\par
\begin{figure}[tbp]
\begin{center}
		\includegraphics[width=8.2cm]{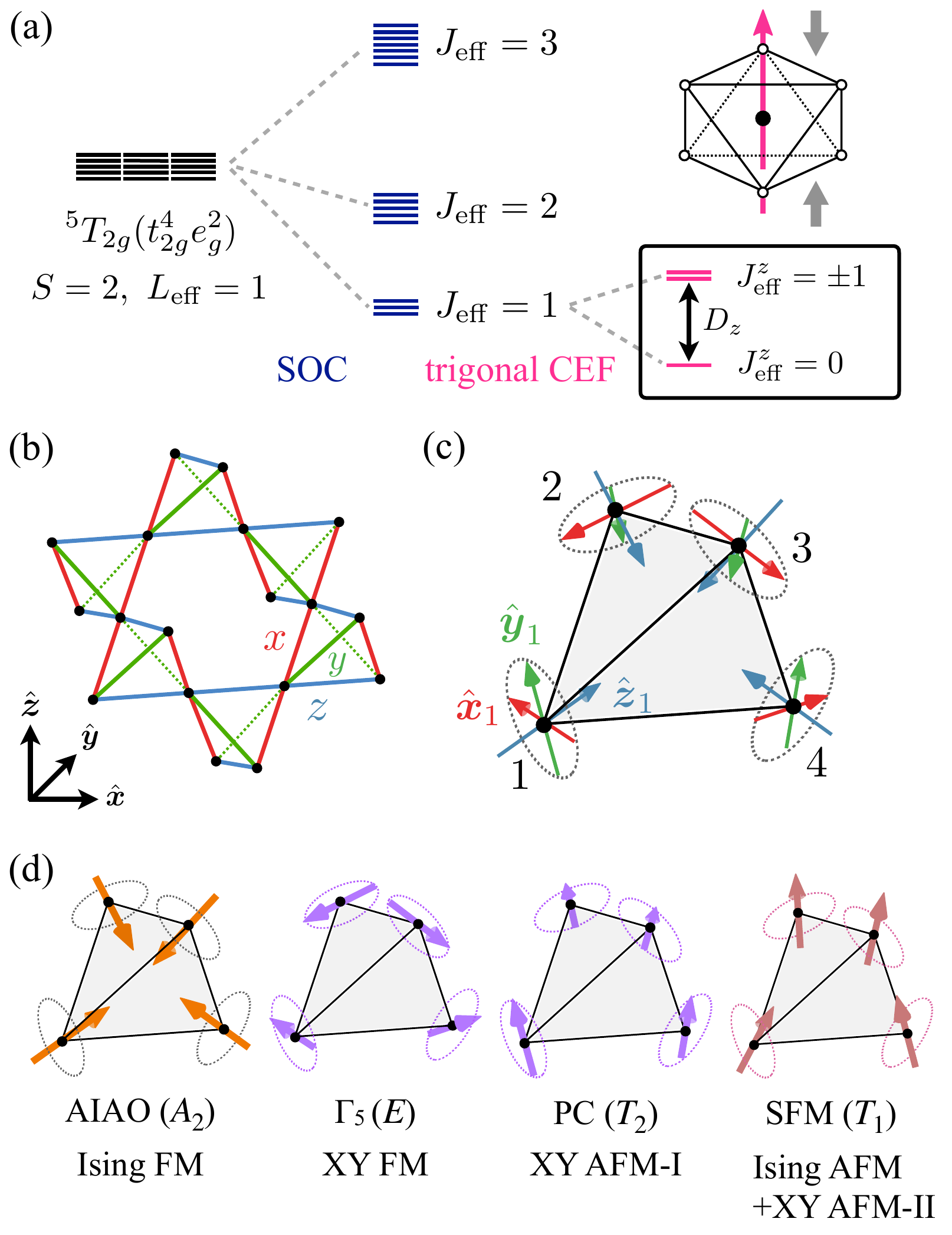}
		\caption{
(a) Atomic multiplet $^5T_{2g} (t^4_{2g}e^2_g)$ of Fe$^{2+}$ ion that split due to SOC and trigonal crystal field. 
(b) Pyrochlore lattice. The $x$, $y$, and $z$ bonds are expressed by red, green, and blue bonds, respectively. The global frame are defined by $\{\hat{\vb*{x}}, \hat{\vb*{y}}, \hat{\vb*{z}}\}$. 
(c) Four sublattices and sublattice-dependent local frames $\{\hat{\vb*{x}}_\mu, \hat{\vb*{y}}_\mu, \hat{\vb*{z}}_\mu\}$ at each sublattice $\mu$. 
(d) Magnetically ordered phases with uniform amplitudes of magnetic moments, designated by irreps of $T_d$.}
\label{f1}
\end{center}
\end{figure}

In this paper, we take account of the spin-orbit coupling (SOC) and a trigonal crystal field effect onto the multiplet. 
The SOC combines the two momenta into $J_{\rm eff}=1$ pseudospin, 
and the trigonal crystal field splits them into a $J_{\rm eff}^z=\pm 1$ doublet and a $J_{\rm eff}^z=0$ singlet [see Fig.~\ref{f1}(a)]. 
The resultant effective Hamiltonian is based on pseudospin-1 degrees of freedom with the 
single-ion anisotropy (SIA) that favors $J_{\rm eff}^z=0$ singlet, and 
the orbital fluctuation effect is reflected in the pseudospin-1 language as the highly anisotropic spin exchange interactions. 
\par
Previous studies on spin-1 systems on a pyrochlore lattice for Ni$^{2+}$-based fluoride 
revealed complex phase diagrams \cite{GangChen2018}, 
where the magnetic phases suffer intense competition due to anisotropic exchange interactions. 
The only manifestation of having $S=1$ is the small window of quantum paramagnetic (QP) phase. 
This QP is distinguished from the conventional classical paramagnet in that 
its magnetic disorder relies on the quantum quadrupole fluctuation 
induced by the on-site potential. 
Typically, the dipolar order trivially accompanies quadrupole order as a mathematical outcome, 
thus the quadrupoles are physically nontrivial when the dipole is absent \cite{Penc2011}. 
Here, however, we propose a case distinct from these conventions; 
dipoles and quadrupoles selectively concentrate on different sublattices that generate a magnetic amplitude wave. 
The only relevant phenomenon reported so far is the CP$^2$ skyrmion, characterized by 
gradually varying magnetic orientations and amplitudes from core to periphery \cite{Amari2022, Zhang2023}. 
\par
The paper is organized as follows. 
Section~\ref{sec:model} introduces the pseudospin-1 low-energy effective Hamiltonian, 
and outlines its derivation, with its full details provided in the Appendix.
Section~\ref{sec:phase_diagram} presents a mean-field phase diagram in the SU(3) framework that allows the full description of quadrupoles. 
Section~\ref{sec:discussion} discusses the relevance of our results with $\rm GeFe_2O_4$ and $\gamma$-$\rm SiFe_2O_4$ 
using the evaluation of material-based parameters. 
\par

\section{Formulation}\label{sec:model}
\subsection{Model Hamiltonian}\label{sec:exchange}
We consider a minimal model for pseudospin-1 pyrochlore magnets given by 
\begin{equation}\label{eq:Ham}
\begin{split}
\mathcal{H} &= \sum_{\langle i,j \rangle} \Big[  J_{zz} S^z_i S^z_j -J_\pm ( S^+_i S^-_j + S^-_i S^+_j )  \\ 
&\qquad\qquad +J_{\pm\pm} (\gamma_{ij} S^+_i S^+_j +\mathrm{H.c.} ) \\
&\qquad\qquad + J_{z\pm} (\zeta_{ij} [S^z_i S^+_j +S^+_i S^z_j ] +\mathrm{H.c.}) \Big] \\
&\quad+\sum_i D_z (S^z_i)^2,
\end{split}
\end{equation} 
whose exchange interactions include bond-dependent phase factors, 
$\gamma_x=-\zeta_x^*=1$, $\gamma_y=-\zeta_y^*=e^{2\pi i/3}$, and $\gamma_z=-\zeta_z^*=e^{-2\pi i/3}$ for $x$, $y$, and $z$ bonds 
[see Fig.~\ref{f1}(b)]. 
The pseudospin-1 operators are represented in a local frame with $z$ axes pointing 
toward the center of the tetrahedron (See Fig.~\ref{f1}(c)). 
The $D_z >0$ term represents the easy-plane SIA produced by the compression of octahedron ligands into trigonal geometry 
in the spinel compounds. 
\par 
The orbital fluctuation effect discussed in Ref.[\onlinecite{Perversi2018}] 
manifests in Eq.(\ref{eq:Ham}) as bond-dependent ($\{x,y,z\}$) and highly anisotropic interactions $\{J_{zz}, J_{\pm}, J_{\pm\pm}, J_{z\pm}\}$ \cite{GangChen2008,Jackeli2009,Khaliullin2013,Liu2018,Khaliullin2021,GangChen2023}.  
Such interactions are known in rare-earth magnets with strong SOC \cite{Ross2011,Rau2019,Li2015,Maksimov2019}, 
while recently recognized as important also in $3d$ materials based on 
$\rm Co^{2+}$ \cite{Songvilay2020,Kim2022,Halloran2023} and $\rm Fe^{2+}$\cite{Bai2021,Legros2021} 
with relatively weak SOC. 
\par
\subsection{Microscopic origin}\label{sec:origin}
Let us briefly explain the microscopic origin of Eq.(\ref{eq:Ham}). 
We start from a $d^6$ state on an isolated Fe$^{2+}$ ion  
and introduce a single-ion Hamiltonian,
\begin{equation}\label{eq:single-ion}
\mathcal{H}_{\rm atom} = \mathcal{H}_{\rm Coulomb} +\mathcal{H}_{\rm CEF}^{\rm cub} +\mathcal{H}_{\rm SOC} +\mathcal{H}_{\rm CEF}^{\rm tri}, 
\end{equation}
which includes Coulomb interaction, crystalline electric field (CEF), and SOC. 
The CEF term is divided into the contribution from the cubic CEF and the trigonal CEF. 
The energy scales of these terms together compared with the interion hopping term $\mathcal{H}_{\rm kin}$ are
\begin{equation}
\mathcal{H}_{\rm Coulomb} > \mathcal{H}_{\rm CEF}^{\rm cub} \gg \mathcal{H}_{\rm kin}  \gg \mathcal{H}_{\rm SOC} \sim \mathcal{H}_{\rm CEF}^{\rm tri}, 
\end{equation}
where $\mathcal{H}_{\rm Coulomb}$ and $\mathcal{H}_{\rm CEF}^{\rm cub}$ are of $1\rm\, eV$ order,  
$\mathcal{H}_{\rm kin}$ is of $\sim 100\rm\, meV$, 
and $\mathcal{H}_{\rm SOC}$ and $\mathcal{H}_{\rm CEF}^{\rm tri}$ is of the order of $\sim 10\rm\, meV$. 
Here we evaluated the ratio of the two CEFs as $\Delta_{\rm tri}/\Delta_{\rm cub} = 0.015$, 
following Ref.~\cite{Khomskii2016} [see Fig.~\ref{fig:S3}(b) in Appendix~\ref{sec:calculation}].  
\par
Diagonalizing the largest two terms, $\mathcal{H}_{\rm Coulomb} +\mathcal{H}_{\rm CEF}^{\rm cub}$, 
we obtain the ground-state multiplet as $^5T_{2g}$, characterized by $S = 2$ and $L_{\rm eff} = 1$
(see Appendix~\ref{sec:Coulob_cubic}). 
The exchange interactions between the neighboring $^5T_{2g}$ multiplets 
are induced by the interion kinetic term, $\mathcal{H}_{\rm kin}$, which yields  
the Kugel-Khomskii (KK) model, ${\cal H}_{\text{KK}}$, including the spin-orbital exchange terms, 
as we detail in Appendix~\ref{sec:Kugel-Khomskii}, which keeps the SU(2) symmetry of spins. 
\par
The remaining terms are $\mathcal{H}_{\rm SOC}=\lambda \bm L_{\rm eff} \cdot \bm S$ and the trigonal CEF. 
Since $\lambda \simeq 13\, \rm meV$ \cite{Dunn1961,Walisinghe2021} and  
the $\Delta_{\rm tri} \simeq 19 \, \rm meV$ (Appendix \ref{sec:calculation-3}) are of comparable magnitude, 
they should, in principle, be treated on equal footing.  
For convenience, we first consider the splitting induced by the SOC, 
which yields $J_{\rm eff} = 1$ triplet, $J_{\rm eff} = 2$ quintet, and $J_{\rm eff} = 3$ septet, 
as shown in Fig.~\ref{f1}(a). 
Subsequently, the trigonal crystal field splits the lowest-energy $J_{\rm eff} = 1$ triplet into $J^z_{\rm eff} = 0$ singlet and $J^z_{\rm eff} = \pm1$ doublet. 
Because we only need to take care of the lowest $J_{\rm eff} = 1$ level, 
${\cal H}_{\text{KK}}$ is projected onto this subspace as
$\mathcal{H}_{\rm eff} = \mathcal{P}_{J_{\rm eff}=1} \mathcal{H}_{\rm KK} \mathcal{P}_{J_{\rm eff}=1}$, 
which includes anisotropic exchange interactions originating from spin-orbital exchange terms 
and a SIA term arising from the trigonal crystal field.
\par
The pseudospin-1 ($J_{\rm eff} = 1$) basis is defined within the local frame of the trigonal CEF, 
whose quantization axis $z_\mu$ is taken as the trigonal axis, 
which is convenient to clarify the role of SIA. 
These basis states are expressed by the linear combinations of $\ket*{S^z, L^z_{\rm eff}}$ of the $^5T_{2g}$ multiplet as 
\begin{eqnarray}
&& \ket*{J_{\rm eff}^{z_\mu}=\pm 1} =  p_1 \ket*{\pm 2, \mp 1} +p_2\ket*{\pm 1, 0} +p_3 \ket*{0, \pm 1} , \nonumber \\
&& \ket*{J_{\rm eff}^{z_\mu}=0} = q_1 \ket*{+1, -1} +q_2 \ket*{0, 0} +q_1 \ket*{-1, +1},
\end{eqnarray}
where the constants $p_i$ and $q_i$ depend on $\lambda$ and $\Delta_{\rm tri}$. 
When $\Delta_{\rm tri}=0$, we have 
$(p_1,p_2,p_3,q_1,q_2)= (\sqrt{6}, -\sqrt{3},1,\sqrt{3},-2) /\sqrt{10}$. 
\par
Let us add a few remarks. 
There are other terms than those in Eq.(\ref{eq:Ham}), 
which are the on-site quadrupole potentials and quadrupole exchange interactions produced by SOC. 
They are one and two orders of magnitude smaller than the dipole exchange interactions, respectively, 
and are safely discarded \cite{GangChen2023} 
[see Appendix \ref{sec:effective_model} and Eq.(\ref{eq:soexchange2})]. 
We finally find that the energy scales discussed here are much higher than the magnetic transition temperature, $T_{N}=9\rm\, K$ \cite{Perversi2018,Barton2014}. 
This supports the inclusion of SOC and trigonal CEF effect as the emergent anisotropic exchange interactions 
between the pseudospin-1 moments. 
The whole detail of the exchange interactions is presented in Appendix~\ref{sec:derivation}. 
\par
\section{Mean-field phase diagram}\label{sec:phase_diagram}
\subsection{Dipolar and quantum para phases}\label{sec:quantum_para}
To see the overall picture, it is convenient to rewrite Eq.(\ref{eq:Ham}) as 
the sum of the local Hamiltonian on each tetrahedron \cite{HanYan2017}, 
$\mathcal{H}=\sum_{\rm tet}\mathcal{H}_{\rm tet}$, where
\begin{equation}\label{eq:Ham2}
\begin{split}
\mathcal{H}_{\rm tet} &= \frac{1}{2} \Big[ 3J_{zz} m_{A_2}^2  -6J_{\pm} \vb*{m}_{E}^2 +(2J_{\pm}-4J_{\pm\pm}) \vb*{m}_{T_2}^2 \\
&\qquad\ -J_{zz}\vb*{m}_{T_1^{(1)}}^2+(2J_{\pm}+4J_{\pm\pm})\vb*{m}_{T_1^{(2)}}^2 \\
&\qquad\ -8J_{z\pm}\vb*{m}_{T_1^{(1)}} \cdot \vb*{m}_{T_1^{(2)}}\Big] +\frac{1}{2}\sum_{\mu \in \rm tet} D_z (S^z_\mu)^2. 
\end{split}
\end{equation}
Here $\vb*{m}_{\Gamma}$ is the irreducible representations (irreps) of dipolar moments on a tetrahedron, 
$T_d=A_2\oplus E\oplus 2T_1 \oplus T_2$ (see Appendix~\ref{sec:point_group}), each written as a linear combination of the spin operators. 
If all the four sites have full dipole moment $M_\mu=1$ ($\mu=1,2,3,4$), which is equivalent to 
the classical spin state or the SU(2) coherent state, 
the SIA term can be renormalized into the $\vb*{m}_\Gamma$ terms \cite{HanYan2017}. 
\par
Let us first consider the two limiting cases. 
The strong SIA limit, $D_z/J_{\gamma\gamma'}=\infty$, has a QP ground state 
with quenched dipole $M_\mu=0$. 
In the opposite limit, $D_z/J_{\gamma\gamma'}=0$, 
we find the magnetic order maximizing the dipole moments as $M_\mu=1$; 
the ground state consists of lowest energy state of $\mathcal{H}_{\rm tet}$ 
over all  the unit cells within the classical frame \cite{HanYan2017}, 
whose magnetic structures are given by either of the irreps of $T_d$; 
all-in/all-out (AIAO) state, $\Gamma_5$ state, Palmer-Chalker (PC) state, 
and splayed ferromagnet (SFM) corresponding to $A_2$, $E$, $T_2$, and $T_1$, respectively 
[see Fig.~\ref{f1}(d)]. 
\par
In the spin language, we call Ising/XY magnets when the spins point along the local $z$ axis/inside the $xy$-plane,
ferromagnet (FM) when all four spins have the same component, 
i.e., $\ev*{\vb*{S}_1}=\ev*{\vb*{S}_2}=\ev*{\vb*{S}_3}=\ev*{\vb*{S}_4}$, and antiferromagnet (AFM) 
when half of the spins point in the two opposite directions as, e.g., $\ev*{\vb*{S}_1}=\ev*{\vb*{S}_2}=-\ev*{\vb*{S}_3}=-\ev*{\vb*{S}_4}$. 
The AIAO state is the Ising FM, $\Gamma_5$ state is the XY FM with accidental U(1) degeneracy about the local $z$ axis,  
and PC state is the XY AFM-I. 
The SFM is named ferromagnet by convention but is in reality a linear combination of Ising AFM (two-in/two-out) 
and XY AFM-II (the $2\pi/3$-rotated PC state for each sublattice spins about the local $z$ axis). 
\par
The ground-state phase diagram of Eq.(\ref{eq:Ham}) was previously studied by Li and Chen \cite{GangChen2018}, 
where they applied the flavor wave theory and the Luttinger-Tisza method. 
Since these methods can take into account only the two limiting cases, 
magnetically ordered $M_\mu=1$ or QP with $M_\mu=0$; 
the AIAO, $\Gamma_5$, PC, SFM, and QP exclusively appear in the phase diagram. 
\par
\begin{figure*}[t]
	\begin{center}
		\includegraphics[width=17.2cm]{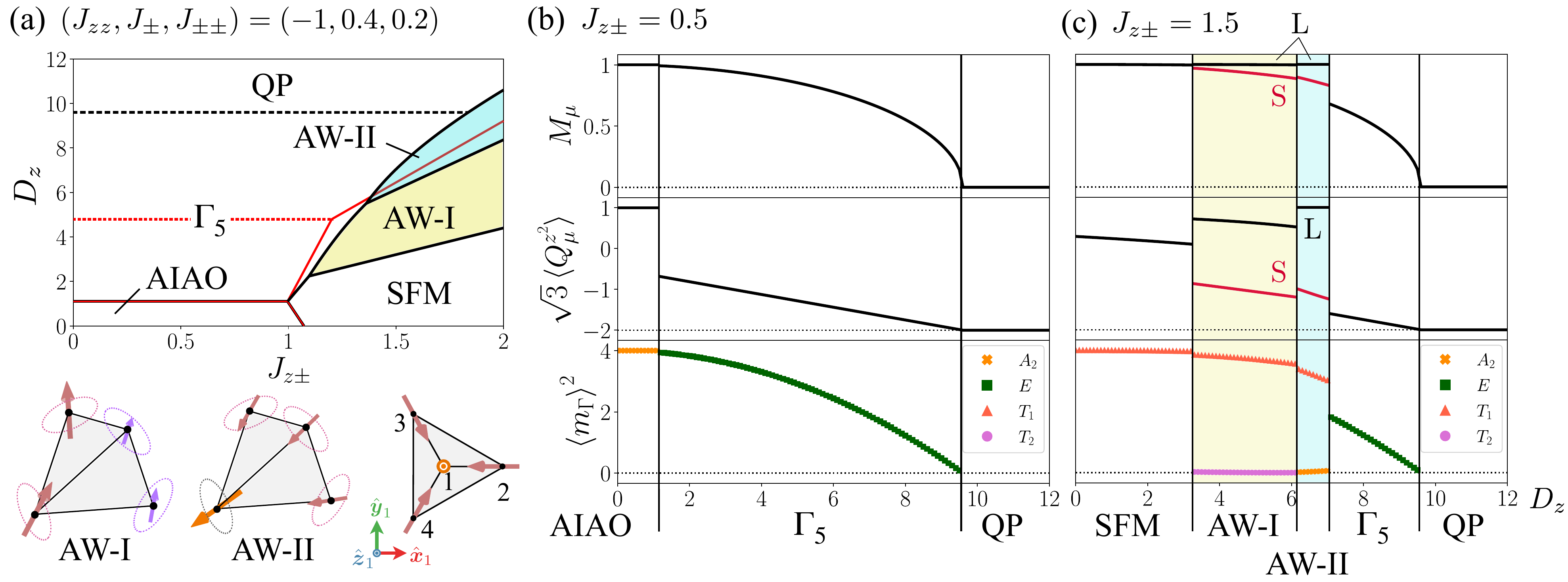}
		\caption{(a) Phase diagram at $(J_{zz}, J_{\pm}, J_{\pm\pm})=(-1, 0.4, 0.2)$ and spin configurations in amplitude-wave-ordered phases. Black (red) lines are phase boundaries obtained by the SU(3) (SU(2)) treatment. Solid (dashed) lines represent discontinuous (continuous) transition. 
			(b)(c) Dipole moments, quadrupole moments, and spin order parameters for (b) $J_{z\pm}=0.5$ and (c) $J_{z\pm}=1.5$. 
			The dipole moment of sublattice $\mu\, (=1,2,3,4)$ is denoted by $M_\mu$, the quadrupole moment by $\ev*{Q^{z^2}_\mu}$, and the spin order parameter by $\ev*{m_\Gamma}^2$, where $\Gamma$ labels irreducible representations of the point group $T_d$, including $A_2$, $E$, $T_1$ , and $T_2$. These moments are separated into large(L) and small(S) moments, 2:2 and 1:3, in AW-I and AW-II phases, respectively. The spin order parameter is plotted when its value is nonzero.
		}
		\label{f2}
	\end{center}
\end{figure*}
\par
\subsection{SU(3) coherent state}\label{sec:su3}
We go beyond the previous approximation by describing the ground state as 
$\ket*{\Psi}=\otimes_j\ket*{\psi_j}$, 
consisting of SU(3) coherent state \cite{Gitman1993,Nemoto2000} on site $j$. 
The SU(2) coherent state is described by two parameters representing the zenithal and azimuthal angles of the dipole vectors; 
\begin{equation}
\begin{split}
\ket*{\psi} &= \frac{1+\cos\theta}{2} e^{-i\varphi} \ket*{+1} \\
&\quad +\frac{\sin\theta}{\sqrt{2}} \ket*{0} +\frac{1-\cos\theta}{2} e^{i\varphi} \ket*{-1}. 
\end{split}
\end{equation}
where $\ket*{\pm 1}$ and $\ket*{0}$ are eigenstates of $S^z$. 
On the other hand, the SU(3) coherent state has four parameters $\{\xi_1, \xi_2, \xi_3, \xi_4\}$;
\begin{equation}
\begin{split}
\ket*{\psi} &= e^{i\xi_3} \cos\xi_2\sin\xi_1 \ket*{+1} \\
&\quad +\cos\xi_1 \ket*{0} +e^{i\xi_4}\sin\xi_2\sin\xi_1 \ket*{-1}, 
\end{split}
\end{equation}
whose eight generators describe the full spin-1 space 
consisting of dipolar $\{S^\alpha\}_{\alpha=x,y,z}$ and quadrupolar $\{Q^{\gamma}\}_{\gamma=3z^2-r^2,{x^2-y^2},xy,yz,zx}$ moments: 
\begin{equation}
\begin{split}
& Q^{3z^2-r^2} = \frac{1}{\sqrt{3}} \Big[2(S^z)^2-(S^x)^2-(S^y)^2 \Big], \\
& Q^{x^2-y^2} = (S^x)^2 -(S^y)^2, \\
& Q^{xy} = S^x S^y +S^y S^x, \\
& Q^{yz} = S^y S^z +S^z S^y, \\
& Q^{zx} = S^z S^x +S^x S^z. 
\end{split}
\end{equation}

This allows us to treat the dipolar-quadrupolar hybridization naturally. 
The four parameters are further reduced by one (see Appendix \ref{sec:spin-1}). 
The three independent parameters are chosen as $\{M_j,\theta_j,\varphi_j\}$ and determined variationally. 
The expectation values of local physical operators are given as
\begin{eqnarray}
\label{eq:spin}&& \ev*{\vb*{S}_j}=M_j(\sin\theta_j\cos\varphi_j,\, \sin\theta_j\sin\varphi_j,\, \cos\theta_j), \\
\label{eq:SIA}&& \ev*{ (S^z_j)^2 } = 1-\frac{1}{2} \sin^2\theta_j \Big(1+\sqrt{1-M_j^2}\Big). 
\end{eqnarray}
Our treatment is in contrast to the SU(2) state designated 
by the angles $(\theta_j,\varphi_j)$ of the fixed $M_j=1$ moment.
\par
\subsection{SU(3) mean-field phase diagram}\label{sec:su3_phase_diagram}
Within the SU(3) framework, we consider the intermediate SIA region, $0<D_z/J_{\gamma\gamma'}<\infty$, 
where $M_j$ is expected to be suppressed toward zero 
when increasing $D_z$ to the critical value, $D_{z,c}$. 
Figure~\ref{f2}(a) shows the phase diagram on the plane of $J_{z\pm}$ and $D_z$ 
with the other parameters fixed to $(J_{zz}, J_{\pm}, J_{\pm\pm})=(-1, 0.4, 0.2)$.
The local $\pi$ rotations about $\hat{\vb*{z}}_\mu$ axes converts 
the sign of $J_{z\pm}$ while keep the other four parameters unchanged, 
and thus we only need to consider $J_{z\pm}\geq 0$. 
In the small-$J_{z\pm}$ region, 
AIAO and $\Gamma_5$ are observed with the same $M_\mu$ for all the four sublattices. 
The order parameters exclusively take single irreps $\ev*{\vb*{m}_{\Gamma}}$ 
and the $\Gamma_5$-to-QP transition at $D_{z,c}$ is continuous, 
as shown in Fig.~\ref{f2}(b) for $J_{z\pm}=0.5$. 
\par
At larger $J_{z\pm}$, in addition to the SFM, 
there are two magnetic phases 
having inequivalent distribution of $\langle M_\mu\rangle$'s among sublattices 
$\mu=1,2,3,4$, which we denote as amplitude wave (AW). 
The fragments of large and small moments appear in a 2:2 ratio among the four sublattices for AW-I, 
whereas for AW-II we find 1:3, which is explicitly shown in Fig.~\ref{f2}(c) for $J_{z\pm}=1.5$. 
The quadrupoles also show the same types of spatially inequivalent distributions. 
This kind of distribution is realized by 
mixing the components of order parameters belonging to irreducible representation;  
$\ev*{\vb*{m}_{T_1}}$ and $\ev*{\vb*{m}_{T_2}}$ for the AW-I, 
and $\ev*{\vb*{m}_{T_1}}$ and $\ev*{m_{A_2}}$ for the AW-II phase.
The quadrupoles are intrinsic to AWs, because if we apply an SU(2) treatment, 
the phase boundaries shift to smaller $D_z$ and the AW phases disappear [see Fig.~\ref{f2}(a)]. 
\par
There are three different types of transitions 
from the magnetic to the QP phase in increasing $D_z$; 
(i) discontinuous transition from AIAO or SFM phases with Ising moment ($\langle S^z\rangle \ne 0$), 
(ii) continuous transition from the XY magnetic phases ($\Gamma_5$ and PC), 
and, in addition, 
(iii) discontinuous transition from the AW phases.
For case (ii) we can analytically show the critical points as, 
$D_{z,c}=24J_{\pm}$ and $D_{z,c}=-8J_{\pm}+16J_{\pm\pm}$ for the $\Gamma_5$-QP and PC-QP transition, 
where the magnetic moments uniformly and continuously decrease toward zero. 
The reason why (i) and (iii) have discontinuity is that 
due to the SIA [see Eq.(\ref{eq:SIA})], there are finite quenched moments in the $z$ direction 
which has no energy gain about shrinking the size of the moment 
with increasing $D_z$, and thus they remain finite until the system is replaced by a QP with zero moment. 
\par

\subsection{Amplitude-wave magnets}\label{sec:aw_magnets}
\begin{figure}[t]
\begin{center}
\includegraphics[width=8cm]{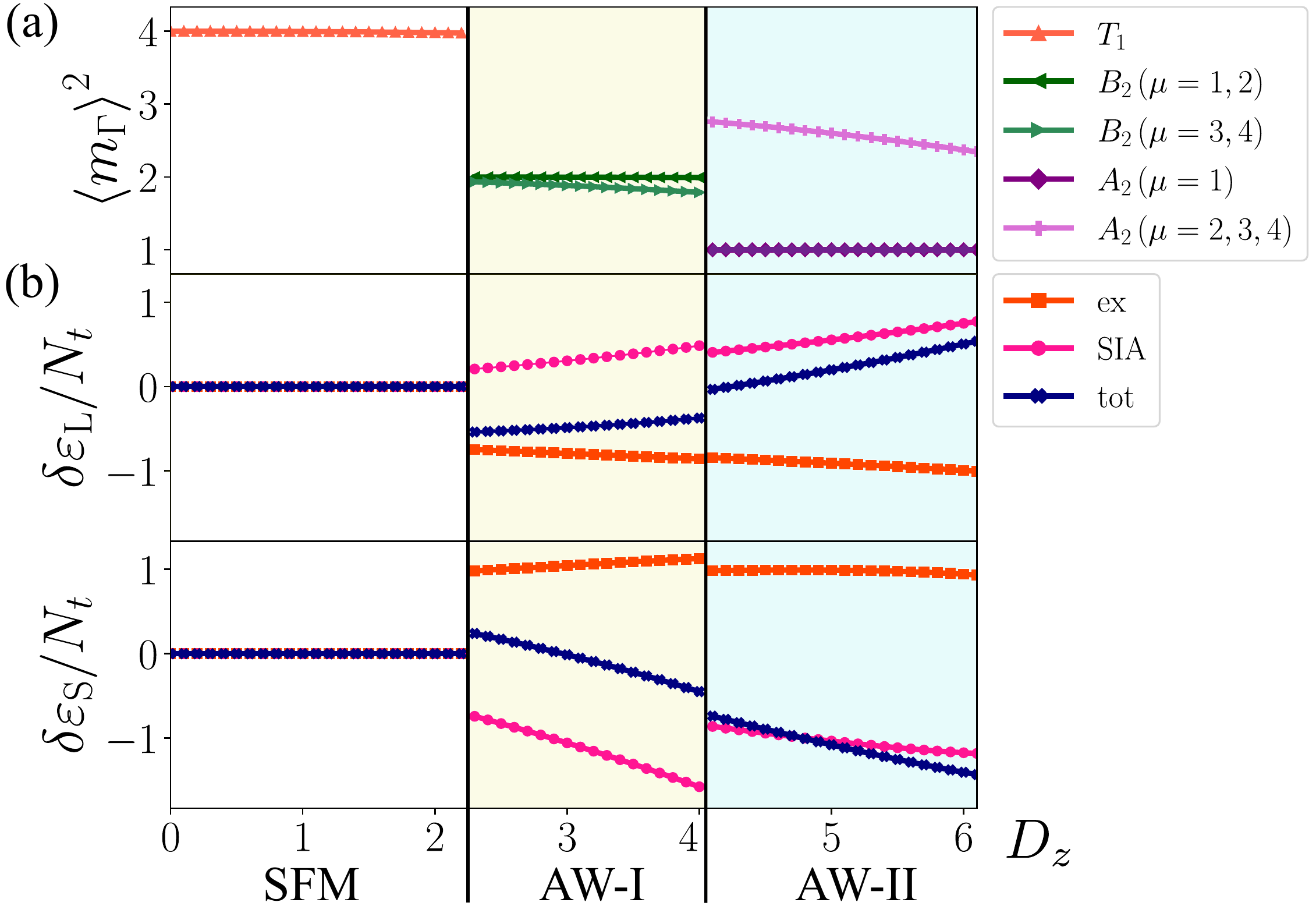}
\caption{
(a) Spin order parameters $\ev*{\vb*{m}_{\Gamma}}$ as a function of $D_z$, 
plotted only when their values are nonzero. 
We plotted the square value, $\ev*{\vb*{m}_{\Gamma}}^2$. 
Because of their time-reversal partners, $\pm \langle m_\Gamma \rangle$, 
their signs may vary at each calculation. 
In each phase (SFM, AW-I, and AW-II), we plot the spin order parameter that belongs to an irreducible representation of the corresponding point group: $T_d$ for the SFM phase, $C_{3v}$ for the AW-I phase, and $C_{2v}$ for the AW-II phase. The explicit forms of these order parameters are listed in Table~\ref{table:spin} of Appendix~\ref{sec:point_group}.
(b) Relative energies of the exchange interaction, SIA, and the total energy per tetrahedron as functions of $D_z$. The quantities for large and small spins are denoted by $\delta\varepsilon_{\rm L}$ and $\delta\varepsilon_{\rm S}$, respectively. The total number of tetrahedra is denoted by $N_t$.
}
\label{f3}
\end{center}
\end{figure}

The symmetry breaking inside the tetrahedron 
requires different treatment of AWs from other magnetically ordered phases; 
we apply the point groups $C_{2v}$ and $C_{3v}$ for the 2:2 and 1:3 fragmentations, respectively. 
In the former case, the irreducible decomposition is $C_{2v}=A_1 \oplus 2A_2\oplus B_1 \oplus 2B_2$ 
for $\mu=1,2$, and $C_{2v}=A_1 \oplus2 A_2\oplus 2B_1 \oplus B_2$ for $\mu=3,4$.
In the latter case, $C_{3v}=A_2\oplus E$ for $\mu=1$ and $C_{3v}=A_1\oplus 2A_2\oplus 3E$ for $\mu=2,3,4$ (see Appendix \ref{sec:point_group}).
Indeed, as shown in Fig.~\ref{f3}(b), 
the order parameters of the same irreps have nonzero values; 
in the AW-I phase, we find $B_2$ with 
\begin{equation}
\begin{split}
& \ev*{\vb*{S}_1}=-\ev*{\vb*{S}_2}= M_{\rm L} (\sin\alpha,\, 0,\, \cos\alpha), \\
& \ev*{\vb*{S}_3}=-\ev*{\vb*{S}_4}= M_{\rm S} (0,\, 1,\, 0), \\
\end{split}
\end{equation}
and for AW-II phase, $A_2$ with 
\begin{equation}\label{eq:AMM1}
\begin{split}
& \ev*{\vb*{S}_1}= M_{\rm L} (0,\, 0,\, 1), \\
& \ev*{\vb*{S}_2}= M_{\rm S} (\sin\alpha,\, 0,\, \cos\alpha), \\
& \ev*{\vb*{S}_3}= M_{\rm S} (-\sin\alpha/2,\, \sqrt{3} \sin\alpha/2,\, \cos\alpha), \\
& \ev*{\vb*{S}_4}= M_{\rm S} (-\sin\alpha/2,\, -\sqrt{3} \sin\alpha/2,\, \cos\alpha), 
\end{split}
\end{equation}
where $M_{\rm L}=1\geq M_{\rm S}$. 
\par
To understand the origin of the AW states, 
we examine the effect of $J_{z\pm}$ and $D_z$, while setting $J_{zz}=J_{\pm}=J_{\pm\pm}=0$ 
because these parameters are irrelevant to AWs. 
The variation of order parameters is shown in Fig.~\ref{f3}(a). 
Without the SIA, i.e., $D_z=0$, the ground state is the SFM whose spins have the same $xy$ and $z$ amplitudes. 
When $D_z\ne 0$, there arise two types of incentives to gain the energy; 
to make the dipole moment large to activate $J_{z\pm}$, 
and to suppress the dipole moment to prevent raising the energy due to SIA. 
The breaking of the equivalence of the four sublattice spins would 
be efficient in coping with this conflict. 
To illustrate it, we write down the mean-field Hamiltonian in a unit of a tetrahedron, 
\begin{equation}\label{eq:Ham3}
\mathcal{H}_{\rm tet} = \frac{1}{2}\sum_{\mu\in\rm tet} \Big[ -\vb*{S}_{\mu} \cdot\vb*{H}'_\mu +D_z (S_\mu^z)^2 \Big] 
+ \text{const.}, 
\end{equation}
where $\vb*{H}'_\mu$ is a sublattice-dependent mean field, e.g., 
\begin{equation}
\vb*{H}'_{1} = J_{z\pm}\mqty( 2S_2^z -S_3^z -S_4^z \\ \sqrt{3}S_3^z -\sqrt{3}S_4^z \\ 2S_2^x -S_3^x+\sqrt{3}S_3^y -S_4^x -\sqrt{3}S_4^y), 
\end{equation}
which equivalently describes our SU(3) product-state solutions. 
The energies of the mean-field term and SIA term on sublattices $\mu$ are separately evaluated. 
As the SFM has both energies in common with all $\mu$, 
we evaluate the energy of AWs from the SFM ones and denote them as 
$\delta \epsilon_{{\rm ex},\mu}$ and $\delta \epsilon_{{\rm SIA},\mu}$. 
Figure~\ref{f3}(b) shows these energies for large and small spins, 
$\delta \epsilon_{\rm ex/SIA, L} =\sum_{\mu \in \rm large} \delta \epsilon_{{\rm ex/SIA}, \mu}$ and 
$\delta \epsilon_{\rm ex/SIA, S} =\sum_{\mu \in \rm small} \delta \epsilon_{{\rm ex/SIA}, \mu}$, and the total energy, 
$\delta \epsilon_{\rm tot,L/S}$ as functions of $D_z$. 
In both phases, the large spins contribute to the exchange interaction despite their disadvantage 
in terms of the SIA, $\delta \epsilon_{\rm ex,L}<0$ and $\delta \epsilon_{\rm SIA,L}>0$, and oppositely for the small spins, 
$\delta \epsilon_{\rm ex,S}>0$ and $\delta \epsilon_{\rm SIA,S}<0$. 
With increasing $D_z$, $M_\mu$ decreases due to SIA, 
while in the AW-I and AW-II states, there are two and one maximized moments. 
With these moments, $J_{z,\pm}M_{\rm L}$ acts as a transverse field on S-sublattices, 
and the S-sublattices take the role to minimize the loss from the SIA by suppressing their moments. 
Why the fragments change from 2:2 to 1:3 is that the importance of the latter effect by the S-sublattices 
become substantial for larger $D_z$.

\begin{table*}[tbp]
\caption{Exchange paraneters $\{J_{zz}, J_{\pm}, J_{\pm\pm}, J_{z\pm}\}$ 
in the local frame and $\{J, K, \Gamma, D\}$ in the global frame 
calculated for $\rm GeFe_2O_4$ and $\gamma$-$\rm SiFe_2O_4$. 
The values for oxygen-mediated hopping only ($x=0$), and direct hopping only ($x=1$) are shown. 
The exchange parameters $\{J, K, \Gamma, \Gamma' \}$ for Kitaev honeycomb materials are also listed for comparison. 
}
	\label{table:local}
		\begin{ruledtabular}
		\begin{tabular}{lrrrrlrrrrr}
		Material  & $J_{zz}$ & $J_{\pm}$ & $J_{\pm\pm}$ & $J_{z\pm}$  &[meV]& $J$ & $K$ & $\Gamma$ & $D \, (\Gamma')$ & [meV] \\ \hline
		${\rm GeFe_2O_4}\ (x=0)$ & $-6.2$   & $5.7$ & $14$  & $16$   &&  $34$   &  $4.7$ & $-5.5$  &   $0.24$  \\
		${\rm GeFe_2O_4}\ (x=1)$ & $-2.4$ &  $-0.66$  &  $0.83$  &  $4.1$  && $4.4$ &  $4.6$  &  $4.0$  &  $-0.23$   \\
		${\gamma\mathchar`-\rm SiFe_2O_4} \ (x=0)$ & $-5.3$ & $7.7$ & $19$  & $16$ && $39$ &  $3.3$ & $-14$ & $2.5$   \\
		${\gamma\mathchar`-\rm SiFe_2O_4} \ (x=1)$ &  $-1.7$  & $-0.67$  & $0.91$ & $3.9$ && $3.9$ & $4.9$ & $3.4$ &  $-0.30$ \\ \hline
		$\rm Na_3Co_2SbO_6$ \cite{Songvilay2020} & $-5.9$ & $2.9$ & $-1.1$ & $-2.4$ && $-2.0$& $-9.0$ & $0.3$ & $-0.8$  \\
		$\rm Na_2Co_2TeO_6$ \cite{Songvilay2020} & $-1.5$ & $2.0$  & $-1.0$  & $-2.5$ && $-0.1$ & $-9.0$ & $1.8$ & $0.3$ \\
		$\alpha\mathchar`-\rm RuCl_3$ \cite{Winter2016} &  $0.4$  & $3.0$  & $7.1$ & $-3.3$ && $-2.2$ & $-5.0$ & $8.0$ & $-0.1$\\ 			 	$\rm Na_2IrO_3$ \cite{Winter2016}  & $-6.8$  & $1.6$ & $-2.4$ & $-4.6$ && $1.6$   & $-17.9$    & $-0.1$  & $-1.8$   
		\end{tabular}
		\end{ruledtabular}
\end{table*}

\section{Summary and Discussion}\label{sec:discussion}

We analyzed the pseudospin-1 effective Hamiltonian with strong single-ion anisotropy (SIA) and 
anisotropic spin exchange interactions, and found the amplitude wave (AW) phases based on the dipole-quadrupole hybridization. 
The model features the spinel oxides such as $\rm GeFe_2O_4$ and $\gamma$-$\rm SiFe_2O_4$, 
whose $\rm Fe^{2+}$ ions form a pyrochlore lattice. 
The magnetic moment, originating from the $S=2$ and $L_{\rm eff}=1$ multiplet $^5T_{2g} (t_{2g}^4 e_g^2)$, 
exhibits spatial modulation in its amplitude, 
an observation that cannot be simply understood by any of the previously established scenarios. 
The key ingredient in our theoretical framework is the inclusion of spin-orbit coupling (SOC) 
and trigonal crystal field effects for $\rm Fe^{2+}$. 
Although these effects are by an order of magnitude smaller than the energy scale of the $^5T_{2g}$ multiplet, 
they remain significantly larger than the N\'eel temperature ($T_N \sim 9$ K) 
and play an essential role in the low-energy physics. 
The SOC hybridizes the spin and orbital components to form a pseudospin-1 
and induces strongly anisotropic spin exchange interactions. 
Meanwhile, the trigonal crystal field lifts the degeneracy of the $J_{\rm eff}^z = 0$ and $\pm1$ levels, 
effectively generating a single-ion anisotropy in the pseudospin-1 space.

To fully capture both dipole and quadrupole degrees of freedom, we employ an SU(3) framework. 
This goes beyond conventional SU(2) descriptions, where magnetic phases are restricted to fully polarized dipoles, 
and quadrupoles only emerge in the quantum paramagnetic phase when SIA lowers their on-site energy. 
In contrast, our SU(3)-based analysis reveals AW phases in which different sublattices of 
the tetrahedron are dominated by either dipolar or quadrupolar moments. 
This is possible because SU(3) allows the moment amplitudes to shrink, 
enabling mixed configurations not accessible in SU(2)-only treatments. 

\par
The major driving force of the AWs is shown to be the two conflicting roles of $J_{z\pm}$, 
i.e., to have a robust $S^z$ dipole on one side and to suppress it on the other side of the bond, 
and the former contributes to the energy gain due to SIA represented by $D_z$. 
To gain the $J_{z\pm}$ term supported by $D_z$, fragmenting the sublattice into two different roles, 
the SFM is converted to AWs by lowering the symmetry. 
As we saw in Fig.~\ref{f1}(d), the SFM can be regarded as the linear combination of 
two-in/two-out Ising AFM and the XY AFM II. 
The introduction of $D_z$ will vary how these two states mix, which can realize both the 2:2 and 1:3 AW configurations. 
\par
We now discuss whether the exchange parameters, particularly $J_{z\pm}$ can take the realistic values in the materials 
to host the AW states 
(for calculation details see Appendix \ref{sec:calculation}).
We consider two types of hopping processes: oxygen-mediated and direct hopping, and parametrize their ratio by $x$ as such that 
$x=0$ corresponds to the case including only the former and $x=1$ the one only with the latter. 
The resultant values of exchange interactions $\{J_{zz}, J_{\pm}, J_{\pm\pm}, J_{z\pm}\}$ are listed in Table~\ref{table:local}. 
We find that for $x=1$, $J_{z\pm}$ becomes the most dominant among the four exchange parameters, 
satisfying the relation $J_{z\pm} > |J_{zz}| > |J_{\pm}|, J_{\pm\pm}$, which is favorable for the realization of AWs. 
\par
These exchange parameters are transformed to the global frame \cite{Rau2018} as
\begin{equation}
\begin{split}
& J=\frac{1}{3} (-J_{zz}+4J_\pm +2J_{\pm\pm} +2\sqrt{2}J_{z\pm}), \\
& K=\frac{2}{3} (J_{zz}-4J_\pm +J_{\pm\pm} +\sqrt{2}J_{z\pm}), \\
& \Gamma=\frac{1}{3}(-J_{zz}-2J_\pm -4J_{\pm\pm} +2\sqrt{2}J_{z\pm}), \\
& D=\frac{\sqrt{2}}{3} (-J_{zz} -2J_\pm +2J_{\pm\pm} -\sqrt{2}J_{z\pm}), 
\end{split}
\label{eq:local-to-global}
\end{equation}
to be compared with other material systems. 
On the right half of Table~\ref{table:local}, we list the transformed values for $\rm GeFe_2O_4$ and $\gamma$-$\rm SiFe_2O_4$ 
together with those of the Kitaev honeycomb magnets. 
For the parameters favorable for AWs ($x=1$), $J$ is suppressed down to the values comparable with $K$ and $\Gamma$. 
This is in sharp contrast to the heavy transition metal magnets with edge-shared octahedra: 
Representative cases of the Kitaev honeycomb magnets \cite{Songvilay2020,Winter2016} are shown in the lower half of the table, 
which has substantially large $K$ and $\Gamma$ compard to $J$. 
Contrastingly, for these cases, the anisotropic exchange, $K$ and $\Gamma$, 
are mediated by the indirect hopping $(x=0)$, 
and the direct hopping generates the isotropic $J$.
\par
Regarding experimentally observed SDW in $\rm GeFe_2O_4$ and $\gamma$-$\rm SiFe_2O_4$, 
our scenario naturally accounts for the bond-dependent exchange energy proposed in Ref.~\cite{Perversi2018} as essential for the formation of SDW; 
they argue the orbital fluctuation of $t_{2g}^4$ in Fe$^{2+}$, 
choosing which of the $t_{2g}$ orbitals extending along the $x,y,z$ bonds 
to be fully occupied and the remaining two to be singly occupied, 
which dynamically determines the active F or AF magnetic interactions. 
This roughly leads to the mixture of two-up/two-down and three-up/one-down configurations per unit tetrahedron. 
Our theory thus gives a more concrete and energetically precise understanding of the formation of AWs, 
the sublattice selective dipolar-quadrupolar hybridization. 
\par
Historically, the SDW state was primarily observed in solids with metallic properties. 
The SDW arises from a metallic state with incommensurate filling factors, 
where a nesting instability induces an SDW with a small moment amplitude, 
preserving metallic properties in two and three dimensions due to imperfect nesting. 
In insulating magnets, spin-1/2 systems generally have limited options to modulate spin moments 
other than forming a supersolid state; the sublattices are fragmented into those with 
diagonal and off-diagonal orders, which alleviates the effects of geometric frustration.
For systems with larger spin values, quantum fluctuations are reduced, 
and one finds either the conventional dipolar magnetic orders with either full moments or, 
at most, uniformly reduced moments due to zero-point fluctuations. 
In our study, however, anisotropic exchange interactions 
naturally present in SOC-based magnets 
play a crucial role in selectively suppressing the magnetic moments in a symmetry-broken manner 
to produce a liquid-solid coexisting AW phase. 

\begin{acknowledgments}
	We thank Karlo Penc for helpful comments. 
	H.N. was supported by a Grant-in-Aid for JSPS Research Fellow (Grant No. JP23KJ0783). 
	This work is supported by KAKENHI Grants No. JP21H05191 and No. 21K03440 
	from JSPS of Japan. 
\end{acknowledgments}

\appendix
\section{Microscopic derivation of effective pseudospin-1 model}\label{sec:derivation}
In this Appendix, we first show how the ground multiplet of $\rm Fe^{2+}$ produced by Coulomb 
interaction and cubic CEF in Sec.~\ref{sec:Coulob_cubic}, 
and then derive the Kugel-Khomskii model which describes spin-orbital exchange interactions between 
the above multiplets in Sec.~\ref{sec:Kugel-Khomskii}. 
Next, we include the effect of SOC and trigonal CEF for the ground multiplet in Sec.~\ref{sec:SOC_trigonal}. 
The resultant low-energy effective Hamiltonian (\ref{eq:Ham}) for pseudospin-1 is obtained in Sec.~\ref{sec:effective_model}. 
The exchange parameters are calculated in Appendix.~\ref{sec:calculation}. 

\begin{widetext}
\subsection{Coulomb interaction and cubic crystalline electric field}\label{sec:Coulob_cubic}

Let us first consider the first and the second terms of atomic Hamiltonian (\ref{eq:single-ion}), whose energy scale is order of $1\rm\, eV$. 
The Coulomb interaction is expressed as
\begin{equation}
\mathcal{H}_{\rm Coulomb} = \frac{1}{2} \sum_{m_1,m_2,m_3,m_4}\sum_{\sigma,\sigma'} \langle m_1m_2||m_3m_4 \rangle c^\dagger_{m_1\sigma} c^\dagger_{m_2\sigma'} c_{m_4\sigma'} c_{m_3\sigma}, 
\end{equation}
where $c^\dagger_{m\sigma}$ creates an electron with orbital $m\in\{d_{yz}, d_{zx}, d_{xy}, d_{3z^2-r^2}, d_{x^2-y^2} \}$ and spin $\sigma\in\{\uparrow,\downarrow\}$. 
The matrix elements of Coulomb interaction are given by
\begin{equation}\label{eq:Coulomb_Coury}
\begin{split}
\langle m_1m_2||m_3m_4 \rangle &= \int\int\dd\vb*{r}\dd\vb*{r}' \phi^*_{m_1}(\vb*{r}) \phi^*_{m_2}(\vb*{r}') \frac{1}{|\vb*{r}-\vb*{r}'|} \phi_{m_3}(\vb*{r}) \phi_{m_4}(\vb*{r}') \\
&= U_{\rm C}\delta_{m_1m_3}\delta_{m_2m_4} +\left( J_{\rm C}+\frac{5}{2}\Delta J_{\rm C}\right) (\delta_{m_1m_4}\delta_{m_2m_3}+\delta_{m_1m_2}\delta_{m_3m_4}) \\
&\quad -48\Delta J_{\rm C} \Tr[\Xi_{m_1}\Xi_{m_2}\Xi_{m_3}\Xi_{m_4}], 
\end{split}
\end{equation}
where $\Xi_m$ is the orbital-dependent $3\times 3$ traceless matrix given by
\begin{equation}
\begin{split}
& \Xi_{yz} = \frac{1}{2}\mqty(0 & 0 & 0 \\ 0 & 0 & 1 \\ 0 & 1 & 0), \quad \Xi_{zx} = \frac{1}{2}\mqty(0 & 0 & 1 \\ 0 & 0 & 0 \\ 1 & 0 & 0), \quad \Xi_{xy} = \frac{1}{2}\mqty(0 & 1 & 0 \\ 1 & 0 & 0 \\ 0 & 0 & 0), \\
& \Xi_{x^2-y^2} = \frac{1}{2}\mqty(1 & 0 & 0 \\ 0 & -1 & 0 \\ 0 & 0 & 0), \quad \Xi_{3z^2-r^2} =\frac{1}{2\sqrt{3}} \mqty(-1 & 0 & 0 \\ 0 & -1 & 0 \\ 0 & 0 & 2), 
\end{split}
\end{equation}
and $U_{\rm C}$, $J_{\rm C}$, and $\Delta J_{\rm C}$ are Coury parameters for $d$-electron systems \cite{Coury2016}; $U_{\rm C}$ denotes the Coulomb integrals between $t_{2g}$ orbitals, $J_{\rm C}$ denotes the average of the exchange integrals between $e_{g}$ one and $t_{2g}$ one, and $\Delta J_{\rm C}$ denotes the difference between the exchange integrals between $e_{g}$ one and $t_{2g}$ one.  
They are related to Racah parameters $(A,B,C)$ or Slater-Condon parameters $(F^0, F^2, F^4)$ as follows:
\begin{equation}
\begin{split}
& U_{\rm C}= A-2B+C = F^0 -\frac{2}{49} F^2 -\frac{4}{441} F^4, \\
& J_{\rm C} = \frac{1}{2}(7B+2C) = \frac{1}{14} F^2 +\frac{5}{126} F^4, \\
& \Delta J_{\rm C} = B = \frac{1}{49} F^2 -\frac{5}{441} F^4. 
\end{split}
\end{equation}
The relationships with the Kanamori parameters for $t_{2g}$ orbitals, $U_{\rm K}$ (intraorbital interaction), $U'_{\rm K}$ (interorbital interaction), and $J_{\rm K}$ (Hund's coupling), are given as 
\begin{equation}
U_{\rm K} = U_{\rm C} +2J_{\rm C} -\Delta J_{\rm C} , \quad 
U'_{\rm K} = U_{\rm C} , \quad 
J_{\rm K} = J_{\rm C} -\frac{1}{2}\Delta J_{\rm C} . 
\end{equation}
	
Considering only Coulomb interaction, the ground multiplet is described by $S=2$, $L=2$ from the Hund's rule since $\rm Fe^{2+}$ ion has six $3d$ electrons. 
This multiplet has $(2S+1)(2L+1)=25$-fold degeneracy, and the eigenstates are labeled by $S_z$ and $L_z$. 
These states are expressed as a superposition of the states in which an orbital with $l_z=L_z$ is doubly occupied and the remaining four orbitals are occupied by one electron each. 

The cubic CEF Hamiltonian is expressed as
\begin{equation}
\mathcal{H}^{\rm cub}_{\rm CEF} = \frac{\Delta_{\rm cub}}{5} \left( 3\sum_{m\in e_g} n_m -2\sum_{m\in t_{2g}} n_m \right), 
\end{equation}
where $n_m=\sum_\sigma c^\dagger_{m\sigma}c_{m\sigma}$ is a number operator of the orbital $m$.
This CEF splits above 25 states into 15 states ($^5T_{2g}$) and 10 states ($^5E_{g}$), similarly to the single-electron $d$ orbitals split into three orbitals ($t_{2g}$) and two orbitals ($e_g$). 
The $^5T_{2g}$ states are expressed as
\begin{equation}
\begin{split}
& \ket*{S^z, YZ} = \frac{i}{\sqrt{2}} (\ket*{S^z, L^z=+1} +\ket*{S^z, L^z=-1}), \\
& \ket*{S^z, ZX} = -\frac{1}{\sqrt{2}} (\ket*{S^z, L^z=+1} -\ket*{S^z, L^z=-1}), \\
& \ket*{S^z, XY} = -\frac{i}{\sqrt{2}} (\ket*{S^z, L^z=+2} -\ket*{S^z, L^z=-2}). 
\end{split}
\end{equation}
We express the orbital part of six-electron states by $\ket*{YZ}$, $\ket*{ZX}$, and $\ket*{XY}$, which corresponds to the doubly occupied single orbital state $d_{yz}$, $d_{zx}$, and $d_{xy}$, respectively.

\subsection{Kugel-Khomskii Hamiltonian}\label{sec:Kugel-Khomskii}

The exchange interactions between the above $^5T_{2g}$ multiplet are described by the Kugel-Khomskii model $\mathcal{H}_{\rm KK}$. 
Here we derive this model by considering $\mathcal{H}_{\rm kin}$ as a perturbative term for $\mathcal{H}_{\rm Coulomb}+\mathcal{H}_{\rm CEF}^{\rm cub}$ and show its bond-dependent nature by virtue of spatial anisotropy of orbitals. 
To represent the model, the following orbital operators are introduced:  
\begin{equation}
\begin{split}
& T^1=I, \quad T^2=L^x, \quad T^3=L^y, \quad T^4=L^z, \\
& T^5=Q^{3z^2-r^2}=\frac{1}{\sqrt{3}} (3(L^z)^2-2I), \quad T^6=Q^{x^2-y^2}=(L^x)^2 -(L^y)^2, \\
& T^7=Q^{yz}=L^y L^z +L^z L^y, \quad T^8=Q^{zx}=L^z L^x +L^x L^z, \quad T^9=Q^{xy}=L^x L^y +L^y L^x, 
\end{split}
\end{equation}
where
\begin{equation}
\begin{split}
& L^x = i\sum_{S^z} \big(\ketbra*{S^z,ZX}{S^z,XY} -\ketbra*{S^z,XY}{S^z,ZX} \big) \\
& L^y = i\sum_{S^z} \big(\ketbra*{S^z,XY}{S^z,YZ} -\ketbra*{S^z,YZ}{S^z,XY} \big) \\
& L^z = i\sum_{S^z} \big(\ketbra*{S^z,YZ}{S^z,ZX} -\ketbra*{S^z,ZX}{S^z,YZ} \big). 
\end{split}
\end{equation}
The resultant Hamiltonian is represented as
\begin{equation}\label{eq:KK}
\mathcal{H}_{\rm KK} = \sum_{\gamma=x,y,z} \sum_{\langle i,j \rangle_\gamma}  \sum_{\mu\nu} \big( K^\gamma_{\mu\nu} T_i^\mu T_j^\nu + J^\gamma_{\mu\nu} T_i^\mu T_j^\nu \vb*{S}_i\cdot\vb*{S}_j \big). 
\end{equation}

The form of spin-orbital exchange interactions depends on the bond, $\mu=x,y,z$. 
Here we focus on the $z$ bond [see Fig.~\ref{fig:S1}(a)]. 
The interactions on the remaining two bonds are obtained by permutation of $\{x,y,z\}$. 
Let us consider the four-sites (two magnetic ions + two ligand ions) problem for the $z$ bond. 
Figure \ref{fig:S1}(b) shows its geometrical configuration. 
The point group of this cluster is $\mathcal{C}_{2v}=\{ \mathcal{E}, \mathcal{C}_2, \sigma_{v}, \sigma_{v'} \}$, where $\mathcal{E}$ is the identity element, $\mathcal{C}_2$ is the twofold rotation, $\sigma_{v}$ and $\sigma_{v'}$ are the mirror operations; 
\begin{equation}
\mathcal{C}_2: \mqty( x_1\\ y_1 \\ z_1 ) \rightarrow \mqty( -x_4\\ -y_4 \\ z_4 ), \quad \sigma_v: \mqty( x_1\\ y_1 \\ z_1 ) \rightarrow \mqty( -y_4\\ -x_4 \\ z_4 ), \quad \sigma_{v'}: \mqty( x_1\\ y_1 \\ z_1 ) \rightarrow \mqty( y_1\\ x_1 \\ z_1 ). 
\end{equation}

\begin{figure*}[tbp]
	\begin{center}
		\includegraphics[width=16cm]{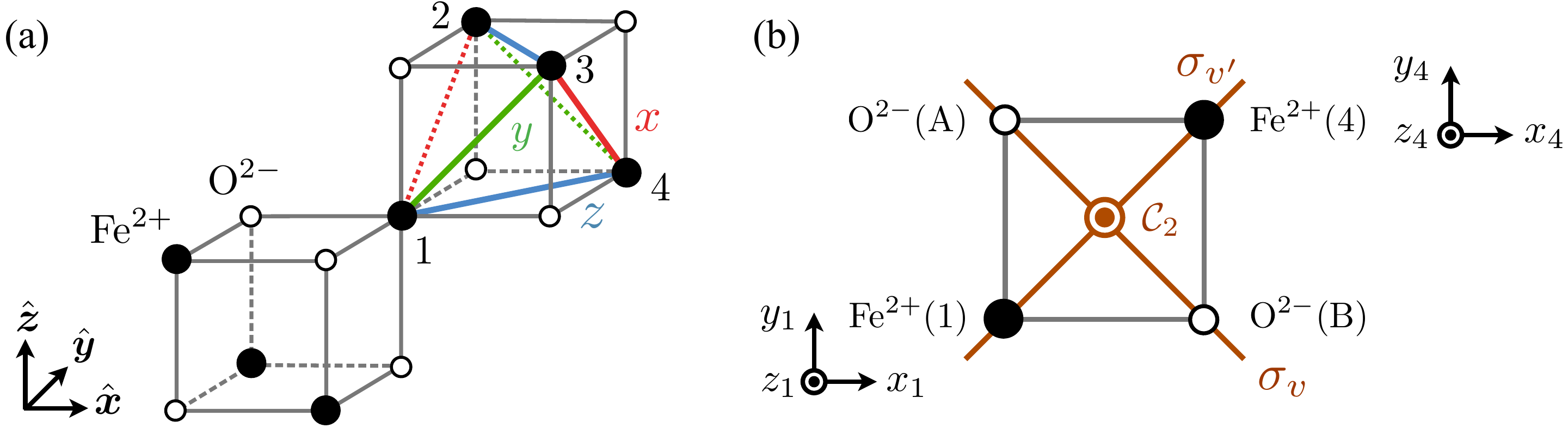}
		\caption{(a) Magnetic ions $\rm Fe^{2+}$ (black) and ligand ions $\rm O^{2-}$ (white) of the spinel $A{\rm Fe_2 O_4}$ ($A={\rm Ge,Si}$). 
			The pyrochlore lattice formed by $\rm Fe^{2+}$ ions has three types of bonds, $x$ (red), $y$ (green), and $z$ bonds (blue). 
			(b) The four-sites cluster consists of two $\rm Fe^{2+}$ ions (1,4) and two ligand ions $\rm O^{2-}$ (A and B). 
			This cluster has three symmetry operations, twofold rotation $\mathcal{C}_2$ and reflections $\sigma_v$, $\sigma_{v'}$.}
		\label{fig:S1}
	\end{center}
\end{figure*}

The perturbation term is the electron hopping between the nearest-neighbor sites: 
\begin{equation}
\mathcal{H}_{\rm kin} = \sum_{\gamma=x,y,z}\sum_{\langle i,j \rangle_\gamma} 
\sum_{mm'}\sum_\sigma t_{ij}^{mm'} c^\dagger_{im\sigma} c_{jm'\sigma}, 
\label{eq:hkin}
\end{equation}
where the hopping matrix is dependent on the type of bonds. 
The hopping process includes the direct hopping between $\rm Fe^{2+}(1)$ and $\rm Fe^{2+}(4)$ and the oxygen-mediated hopping [$\rm O^{2-}(A)$ or $\rm O^{2-}(B)$], as shown in Fig.~\ref{fig:S1}(b). 
Since the virtual electron transfer process does not break SU(2) symmetry in spin space, the Hamiltonian should be expressed as
\begin{equation}\label{eq:KKz}
\mathcal{H}^z_{14} = \sum_{\mu\nu} \big( K^z_{\mu\nu} T_1^\mu T_4^\nu + J^z_{\mu\nu} T_1^\mu T_4^\nu \vb*{S}_1\cdot\vb*{S}_4 \big). 
\end{equation}
The point group $\mathcal{C}_{2v}$ and time-reversal symmetry impose a constraint on the exchange parameters. 
The number of independent parameters is restricted to 18 each in $K$ and $J$. 
The $K$ terms, which do not include the spin operators, are expressed as follows;
\begin{equation}
\begin{split}
\mathcal{H}_{14}^{z, \rm orb} &= A_0 +A_1 (Q_1^{3z^2-r^2} +Q_4^{3z^2-r^2}) +A_2 (Q_1^{xy} +Q_4^{xy}) +A_3(Q_1^{yz}-Q_4^{yz} +Q_1^{zx} -Q_4^{zx}) \\
&\quad +B_\perp (L_1^x L_4^x +L_1^y L_4^y) +B_\parallel L_1^z L_4^z \\
&\quad +B' (L_1^x L_4^y +L_1^y L_4^x) +B'' ( L_1^x L_4^z -L_1^z L_4^x +L_1^y L_4^z -L_1^z L_4^y ) \\
&\quad +C_1 Q_1^{3z^2-r^2} Q_4^{3z^2-r^2} +C_2 Q_1^{x^2-y^2} Q_4^{x^2-y^2} +C_3 (Q_1^{yz} Q_4^{yz} + Q_1^{zx} Q_4^{zx}) +C_4 Q_1^{xy} Q_4^{xy} \\
&\quad +D_1 (Q_1^{3z^2-r^2} Q_4^{xy} +Q_1^{xy} Q_4^{3z^2-r^2}) +D_2 (Q_1^{yz} Q_4^{zx} +Q_1^{zx} Q_4^{yz}) \\
&\quad +E_1 (Q_1^{3z^2-r^2} Q_4^{yz} -Q_1^{yz} Q_4^{3z^2-r^2} +Q_1^{3z^2-r^2} Q_4^{zx} -Q_1^{zx} Q_4^{3z^2-r^2}) \\
&\quad +E_2 (Q_1^{x^2-y^2} Q_4^{yz} -Q_1^{yz} Q_4^{x^2-y^2} -Q_1^{x^2-y^2} Q_4^{zx} +Q_1^{zx} Q_4^{x^2-y^2}) \\
&\quad +E_3 (Q_1^{yz} Q_4^{xy} -Q_1^{xy} Q_4^{yz} +Q_1^{zx} Q_4^{xy} -Q_1^{xy} Q_4^{zx}), 
\end{split}
\end{equation}
where
\begin{equation}
\begin{split}
& A_1=K^z_{15}=K^z_{51}, \quad A_2=K^z_{19}=K^z_{91}, \quad A_3=K^z_{71}=K^z_{81}=-K^z_{17}=-K^z_{18}, \\
& B_\perp =K^z_{22}=K^z_{33}, \quad B_\parallel=K^z_{44}, \quad B'=K^z_{23}=K^z_{32}, \quad B''=K^z_{24}=K^z_{34}=-K^z_{42}=-K^z_{43}, \\
& C_1=K^z_{55}, \quad C_2=K^z_{66}, \quad C_3=K^z_{77}=K^z_{88}, \quad C_4=K^z_{99}, \\
& D_1=K^z_{59}=K^z_{95}, \quad D_2=K^z_{78}=K^z_{87}, \\
& E_1=K^z_{57}=K^z_{58}=-K^z_{75}=-K^z_{85}, \quad E_2=K^z_{67}=-K^z_{68}=-K^z_{76}=K^z_{86},\\
& E_3=K^z_{79}=K^z_{89}=-K^z_{97}=-K^z_{98}. 
\end{split}
\end{equation}
Its matrix representation is as follows;
\begin{equation}
\vb*{T}_1^{\rm T}
\left(\begin{array}{c| c c c| c c c c c}
A_0 & 0 & 0 & 0 & A_1 & 0  & -A_3 & -A_3 & A_2 \\ \hline
0 & B_\perp & B' & B'' & 0 & 0 & 0 & 0 & 0 \\ 
0 & B' & B_\perp & B'' & 0 & 0 & 0 & 0 & 0 \\ 
0 & -B'' & -B'' & B_\parallel & 0 & 0 & 0 & 0 & 0 \\ \hline
A_1& 0 & 0 & 0 & C_1 & 0 & E_1 & E_1 & D_1 \\ 
0 & 0 & 0 & 0 & 0 & C_2 & E_2 & -E_2 & 0  \\ 
A_3 & 0 & 0 & 0 & -E_1 & -E_2 & C_3 & D_2 & E_3 \\ 
A_3 & 0 & 0 & 0 & -E_1 & E_2 & D_2 & C_3 & E_3 \\ 
A_2 & 0 & 0 & 0 & D_1 & 0 & -E_3 & -E_3 & C_4 \\ 
\end{array}\right) \vb*{T}_4, 
\end{equation}
where $\vb*{T}_i^{\rm T}=(T_i^1, T_i^2, \ldots, T_i^9)$. 
Since the same argument applies to $J$ terms including spin operators, 
the coefficient corresponding to $A_1$, for example, is denoted by $\tilde{A}_1$, while for $A_0$ it is denoted by $J$, which expresses the Heisenberg interaction $J\vb*{S}_1\cdot\vb*{S}_4$. 

\subsection{Spin-orbit coupling and trigonal crystalline electric field}\label{sec:SOC_trigonal}
We now consider the SOC and trigonal CEF, which are the third and the fourth terms atomic Hamiltonian (\ref{eq:single-ion}), respectively, 
both having the energy scale of order $10\rm\, meV$. 

We first include the SOC term, which is expressed within the $^5T_{2g}$ manifold as
\begin{equation}
\mathcal{H}_{\rm SOC} = -\lambda \vb*{L}_{\rm eff}\cdot\vb*{S}, \quad \lambda < 0. 
\end{equation}
The effective angular momentum $\vb*{L}_{\rm eff}=-\vb*{L}$ is expressed as
\begin{equation}
L^x_{\rm eff} = \frac{1}{\sqrt{2}} \mqty( 0 & 1 & 0 \\ 1 & 0 & 1 \\ 0 & 1 & 0 ), \quad
L^y_{\rm eff} = \frac{1}{\sqrt{2}} \mqty( 0 & -i & 0 \\ i & 0 & -i \\ 0 & i & 0 ), \quad
L^z_{\rm eff} = \mqty( 1 & 0 & 0 \\ 0 & 0 & 0 \\ 0 & 0 & -1 ), 
\end{equation}
spanned by the eigenstates of $L^z_{\rm eff}$, given by 
\begin{equation}
\begin{split}
& \ket*{L^z_{\rm eff} = \pm 1} = \mp \frac{1}{\sqrt{2}} \ket*{YZ} -\frac{i}{\sqrt{2}} \ket*{ZX}, \\
& \ket*{L^z_{\rm eff} = 0} = \ket*{XY}. 
\end{split}
\end{equation}
Using these states, one can construct a multiplet classified by the effective total angular momentum 
$\vb*{J}_{\rm eff}=\vb*{L}_{\rm eff}+\vb*{S}$; 
the eigenstates of  $\mathcal{H}_{\rm SOC}$ are labeled by $J_{\rm eff}$ and $J^z_{\rm eff}$ because 
$\vb*{J}_{\rm eff}^2$ and $ J^z_{\rm eff}$ commute with $\mathcal{H}_{\rm SOC} = -\lambda \vb*{L}_{\rm eff}\cdot\vb*{S}$. 
The explicit form of the $J_{\rm eff}$ multiplets, $(\ket*{J_{\rm eff}, J^z_{\rm eff}})$, are expressed using $\ket*{S^z, L^z_{\rm eff}}$ as follows:
\begin{itemize}
	\item $J_{\rm eff}=1$ multiplets with the eigenenergy $3\lambda$:
	\begin{equation}
	\begin{split}
	& \ket*{1, \pm 1} = \frac{1}{\sqrt{10}} \big( \sqrt{6}\ket*{\pm 2, \mp 1} -\sqrt{3}\ket*{\pm 1, 0} +\ket*{0, \pm 1} \big), \\
	& \ket*{1, 0} = \frac{1}{\sqrt{10}} \big( \sqrt{3}\ket*{+1, -1} -2\ket*{0, 0} +\sqrt{3} \ket*{-1, +1} \big).
	\end{split}
	\end{equation}
	\item $J_{\rm eff}=2$ multiplets with the eigenenergy $\lambda$:
	\begin{equation}
	\begin{split}
	& \ket*{2, \pm 2} = \frac{1}{\sqrt{3}} \big( \sqrt{2}\ket*{\pm 2, 0} -\ket*{\pm 1, \pm 1} \big), \\
	& \ket*{2, \pm 1} = \frac{1}{\sqrt{6}} \big( \sqrt{2}\ket*{\pm 2,\mp 1} +\ket*{\pm 1, 0} -\sqrt{3} \ket*{0, \pm 1} \big), \\
	& \ket*{2, 0} = \frac{1}{\sqrt{2}} \big( \ket*{+1, -1} - \ket*{-1, +1} \big).
	\end{split}
	\end{equation}
	\item $J_{\rm eff}=3$ multiplets with the eigenenergy $-2\lambda$:
	\begin{equation}
	\begin{split}
	& \ket*{3, \pm 3} = \ket*{\pm 2,\pm 1}, \\
	& \ket*{3, \pm 2} = \frac{1}{\sqrt{3}} \big( \ket*{\pm 2, 0} +\sqrt{2}\ket*{\pm 1, \pm 1} \big), \\
	& \ket*{3, \pm 1} = \frac{1}{\sqrt{15}} \big( \ket*{\pm 2, \mp 1} +2\sqrt{2}\ket*{\pm 1, 0} +\sqrt{6}\ket*{0, \pm 1} \big), \\
	& \ket*{3, 0} = \frac{1}{\sqrt{5}} \big( \ket*{+1, -1} +\sqrt{3}\ket*{0,0} +\ket*{-1, +1} \big).
	\end{split}
	\end{equation}
\end{itemize}

\begin{figure*}[tbp]
	\begin{center}
		\includegraphics[width=18cm]{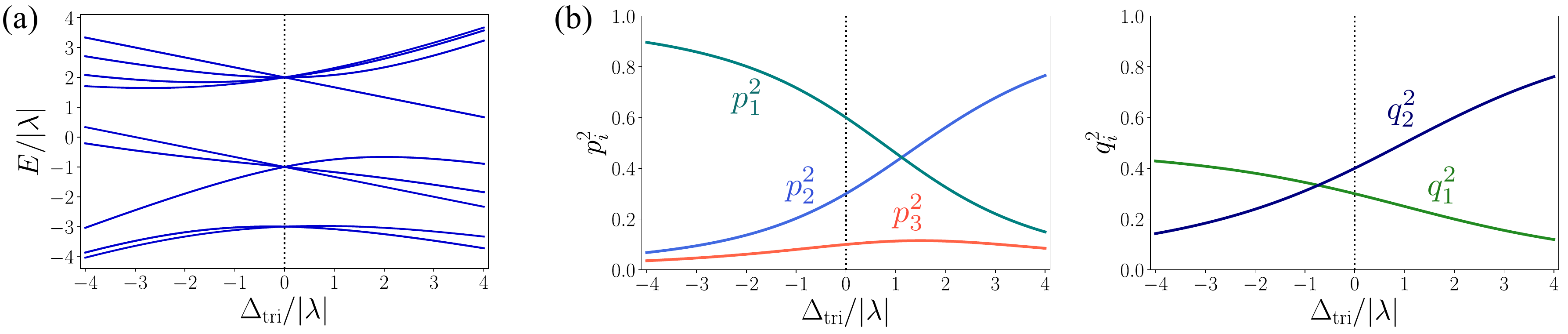}
		\caption{
		(a) Energy spectrum of the Hamiltonian $\mathcal{H}_{\rm SOC} + \mathcal{H}_{{\rm CEF},\mu}^{\rm tri}$ within $^5T_{2g}$ manifold as a function of $\Delta_{\rm tri}/|\lambda|$.
		(b) Weights of the wave function of the $J_{\rm eff}=1$ multiplet multiplet as a function of $\Delta_{\rm tri}/|\lambda|$. The quantities $\{p^2_i\}$ and $\{q^2_i\}$ represent the weights of the components in the states $\ket*{J_{\rm eff}^{z_\mu}=\pm 1}$ and $\ket*{J_{\rm eff}^{z_\mu}=0}$, respectively.}
		\label{fig:S2}
	\end{center}
\end{figure*}

We next introduce the trigonal CEF effect, which 
distorts the octahedron ligand in the direction of the $C_{3v}$-rotational axis $z_\mu$ for the $\rm Fe^{2+}$ site 
belonging to the $\mu$ th sublattice. Such an axis runs in the direction that connects the centers of the two adjacent tetrahedra 
that share the $\rm Fe^{2+}$ site. 
We take the quantization axis as sublattice dependent, whose local frame is introduced as
\begin{equation}
\begin{split}
& \hat{\vb*{x}}_1 = \frac{1}{\sqrt{6}} (-2, +1, +1), \quad \hat{\vb*{y}}_1 = \frac{1}{\sqrt{2}} (0, -1, +1), \quad \hat{\vb*{z}}_1 = \frac{1}{\sqrt{3}} (+1, +1, +1), \\
& \hat{\vb*{x}}_2 = \frac{1}{\sqrt{6}} (-2, -1, -1), \quad \hat{\vb*{y}}_2 = \frac{1}{\sqrt{2}} (0, +1, -1), \quad \hat{\vb*{z}}_2 = \frac{1}{\sqrt{3}} (+1, -1, -1), \\
& \hat{\vb*{x}}_3 = \frac{1}{\sqrt{6}} (+2, +1, -1), \quad \hat{\vb*{y}}_3 = \frac{1}{\sqrt{2}} (0, -1, -1), \quad \hat{\vb*{z}}_3 = \frac{1}{\sqrt{3}} (-1, +1, -1), \\
& \hat{\vb*{x}}_4 = \frac{1}{\sqrt{6}} (+2, -1, +1), \quad \hat{\vb*{y}}_4 = \frac{1}{\sqrt{2}} (0, +1, +1), \quad \hat{\vb*{z}}_4 = \frac{1}{\sqrt{3}} (-1, -1, +1), 
\end{split}
\end{equation}
In this local frame, the representation of the eigenstates $\ket*{L_{\rm eff}^{z_\mu}}$ becomes sublattice dependent; e.g., for $\mu=1$, 
\begin{equation}
\begin{split}
& \ket*{L_{\rm eff}^{z_1}=\pm 1} = \pm \frac{1}{\sqrt{3}} (\ket*{YZ} + e^{\pm 2\pi i/3}\ket*{ZX} +e^{\mp 2\pi i/3}\ket*{XY}), \\
& \ket*{L_{\rm eff}^{z_1}=0} = \frac{1}{\sqrt{3}} (\ket*{YZ} + \ket*{ZX} +\ket*{XY}), 
\end{split}
\end{equation}
using the wave functions $\ket*{YZ}$, $\ket*{ZX}$, and $\ket*{XY}$ in the global frame. 
Within this local basis, $\{ \ket*{L^{z_\mu}_{\rm eff}=+1}, \ket*{L^{z_\mu}_{\rm eff}=0}, \ket*{L^{z_\mu}_{\rm eff}=-1} \}$, 
the trigonal CEF Hamiltonian is expressed as 
\begin{equation}
\mathcal{H}_{{\rm CEF},\mu}^{\rm tri} = \frac{\Delta_{\rm tri}}{3} \mqty( +1 & 0 & 0 \\ 0 & -2 & 0 \\ 0 & 0 & +1). 
\end{equation}
By diagonalizing $\mathcal{H}_{\rm SOC} + \mathcal{H}_{{\rm CEF},\mu}^{\rm tri}$, one finds that the 
$J_{\rm eff}=1$ multiplet splits into the doublet $\ket*{J_{\rm eff}\sim1, J_{\rm eff}^{z_\mu}=\pm 1}$ and the singlet $\ket*{J_{\rm eff}\sim1, J_{\rm eff}^{z_\mu}=0}$, as shown in Fig.~\ref{fig:S2}(a). 
When the octahedron is compressed (elongated) or equivalently, $\Delta_{\rm tri}$ is positive (negative), 
the ground multiplet is the singlet (doublet). 
These states are expressed as
\begin{equation}
\begin{split}
& \ket*{J_{\rm eff}^{z_\mu}=\pm 1} =  p_1 \ket*{\pm 2, \mp 1} +p_2\ket*{\pm 1, 0} +p_3 \ket*{0, \pm 1} , \\
& \ket*{J_{\rm eff}^{z_\mu}=0} = q_1 \ket*{+1, -1} +q_2 \ket*{0, 0} +q_1 \ket*{-1, +1},
\end{split}
\end{equation}
using $\ket*{S^{z_\mu}, L^{z_\mu}_{\rm eff}}$ states. 
The coefficients $p_i$ and $q_i$ are the parameteres dependent on $\lambda$ and $\Delta_{\rm tri}$. 
In the octahedral limit, $p_1=\sqrt{3/5}$, $p_2=-\sqrt{3/10}$, $p_3=1/\sqrt{10}$, $q_1=\sqrt{3/10}$, and $q_2=-2/\sqrt{10}$. 
When we take the fictitiously large limit of the trigonal crystal field, $\Delta_{\rm tri}\rightarrow+\infty$, we find $p_2=q_2=1$, 
and in the opposite limit, $\Delta_{\rm tri}\rightarrow-\infty$, $p_1=q_1=1$. 
Figure \ref{fig:S2}(b) shows these parameters as a function of $\Delta_{\rm tri}/|\lambda|$.
Since the octahedral compression favors $L_{\rm eff}^z=0$, $p^2_2$ and $q^2_2$ increase when $\Delta_{\rm tri}>0$.

\subsection{Low-energy effective Hamiltonian for pseudospin-1 states}\label{sec:effective_model}

Our final goal is to derive the low-energy effective Hamiltonian of the $J_{\rm eff}=1$ subspace. 
This Hamiltonian is obtained by projecting the Kugel-Khomskii Hamiltonian (\ref{eq:KK}) onto the $J_{\rm eff}=1$ manifold, 
$\mathcal{H}_{\rm eff} = \mathcal{P}_{J_{\rm eff}=1} \mathcal{H}_{\rm KK} \mathcal{P}_{J_{\rm eff}=1}$. 

Let us first consider the case without the trigonal crystal field in the global frame. 
This treatment tells us why and to what extent quadrupole interactions are small. 
When projecting onto $J_{\rm eff}=1$ manifold, each operator is represented by $J$ operators \cite{GangChen2023}:
\begin{equation}
\begin{split}
&L^\mu \rightarrow \frac{1}{2} J^\mu, \quad S^\mu \rightarrow \frac{3}{2} J^\mu, \quad Q^\alpha \rightarrow \frac{1}{10} J^{\alpha}, \\
&S^xL^x \rightarrow I -\frac{\sqrt{3}}{10} J^{3z^2-r^2} +\frac{3}{10} J^{x^2-y^2}, \quad S^xL^y \rightarrow \frac{3}{10} J^{xy}, \quad S^xL^z \rightarrow \frac{3}{10} J^{zx}, \\
&S^xQ^{3z^2-r^2} \rightarrow -\frac{\sqrt{3}}{10} J^x, \quad S^xQ^{x^2-y^2} \rightarrow \frac{3}{10} J^x, \\
&S^xQ^{yz}\rightarrow 0, \quad S^xQ^{zx} \rightarrow \frac{3}{10} J^z, \quad S^xQ^{xy} \rightarrow \frac{3}{10} J^y,\\
&S^yL^x \rightarrow \frac{3}{10} J^{xy}, \quad S^yL^y \rightarrow I-\frac{\sqrt{3}}{10}J^{3z^2-r^2}-\frac{3}{10} J^{x^2-y^2}, \quad S^yL^z \rightarrow \frac{3}{10} J^{yz}, \\
&S^yQ^{3z^2-r^2} \rightarrow -\frac{\sqrt{3}}{10} J^y, \quad S^yQ^{x^2-y^2} \rightarrow -\frac{3}{10} J^y, \\
&S^yQ^{yz}\rightarrow\frac{3}{10} J^z, \quad S^yQ^{zx} \rightarrow 0, \quad S^yQ^{xy} \rightarrow \frac{3}{10} J^x, \\
&S^zL^x \rightarrow \frac{3}{10} J^{zx}, \quad S^zL^y \rightarrow \frac{3}{10} J^{yz}, \quad S^zL^z \rightarrow I+\frac{\sqrt{3}}{5} J^{3z^2-r^2}, \\
&S^zQ^{3z^2-r^2} \rightarrow \frac{\sqrt{3}}{5} J^z, \quad S^zQ^{x^2-y^2} \rightarrow 0, \\
&S^zQ^{yz}\rightarrow\frac{3}{10}J^y, \quad S^zQ^{zx} \rightarrow \frac{3}{10}J^x, \quad S^zQ^{xy} \rightarrow 0, 
\end{split}
\end{equation}
where
\begin{equation}
\begin{split}
&J^x = \frac{1}{\sqrt{2}} \mqty( 0 & 1 & 0 \\ 1 & 0 & 1 \\ 0 & 1 & 0 ), \quad
J^y = \frac{1}{\sqrt{2}} \mqty( 0 & -i & 0 \\ i & 0 & -i \\ 0 & i & 0 ), \quad
J^z = \mqty( 1 & 0 & 0 \\ 0 & 0 & 0 \\ 0 & 0 & -1 ), \\
&J^{3z^2-r^2}=\frac{1}{\sqrt{3}} (3(J^z)^2-2I), \quad J^{x^2-y^2}=(J^x)^2 -(J^y)^2, \\
&J^{yz}=J^y J^z +J^z J^y, \quad J^{zx}=J^z J^x +J^x J^z, \quad J^{xy}=J^x J^y +J^y J^x, 
\end{split}
\end{equation}
in the basis of $\{ \ket*{J_{\rm eff}=1, J_{\rm eff}^z=+1}, \ket*{J_{\rm eff}=1, J_{\rm eff}^z=0}, \ket*{J_{\rm eff}=1, J_{\rm eff}^z=-1} \}$. 
In particular, $Q^\alpha$ is mapped to the corresponding $J$ operator with one order of smaller magnitude than $L^\mu$ or $S^\mu$. 
The spin-orbital exchange interaction for $z$ bond (\ref{eq:KKz}) is mapped to
\begin{equation}
\begin{split}
\mathcal{H}_{14}^{z,\rm eff} &= \mathcal{A}_1 (J_1^{3z^2-r^2} +J_4^{3z^2-r^2}) +\mathcal{A}_2 (J_1^{xy}+J_4^{xy}) +\mathcal{A}_3(J_1^{yz} -J_4^{yz} +J_1^{zx} -J_4^{zx}) \\
&\quad +\mathcal{B}_\perp (J_1^x J_4^x +J_1^y J_4^y) +\mathcal{B}_\parallel J_1^z J_4^z \\
&\quad +\mathcal{B}' (J_1^x J_4^y +J_1^y J_4^x) +\mathcal{B}'' ( J_1^x J_4^z -J_1^z J_4^x +J_1^y J_4^z -J_1^z J_4^y ) \\
&\quad +\mathcal{C}_1 J_1^{3z^2-r^2} J_4^{3z^2-r^2} +\mathcal{C}_2 J_1^{x^2-y^2} J_4^{x^2-y^2} +\mathcal{C}_3 (J_1^{yz} J_4^{yz} + J_1^{zx} J_4^{zx}) +\mathcal{C}_4 J_1^{xy} J_4^{xy} \\
&\quad +\mathcal{D}_1 (J_1^{3z^2-r^2} J_4^{xy} +J_1^{xy} J_4^{3z^2-r^2}) +\mathcal{D}_2 (J_1^{yz} J_4^{zx} +J_1^{zx} J_4^{yz}) \\
&\quad +\mathcal{E}_1 (J_1^{3z^2-r^2} J_4^{yz} -J_1^{yz} J_4^{3z^2-r^2} +J_1^{3z^2-r^2} J_4^{zx} -J_1^{zx} J_4^{3z^2-r^2}) \\
&\quad +\mathcal{E}_2 (J_1^{x^2-y^2} J_4^{yz} -J_1^{yz} J_4^{x^2-y^2} -J_1^{x^2-y^2} J_4^{zx} +J_1^{zx} J_4^{x^2-y^2}) \\
&\quad +\mathcal{E}_3 (J_1^{yz} J_4^{xy} -J_1^{xy} J_4^{yz} +J_1^{zx} J_4^{xy} -J_1^{xy} J_4^{zx}), 
\end{split}
\label{eq:soexchange2}
\end{equation}
where
\begin{align}
& \mathcal{A}_1 = \frac{1}{10} A_1-\frac{\sqrt{3}}{5}\tilde{B}_\perp +\frac{\sqrt{3}}{5}\tilde{B}_\parallel,\rule{5mm}{0mm}
 \mathcal{A}_2 = \frac{1}{10} A_2 +\frac{3}{5}\tilde{B}',\rule{5mm}{0mm}
 \mathcal{A}_3 = \frac{1}{10} A_3,\notag \\
& \mathcal{B}_\perp = \frac{1}{4}B_\perp +\frac{9}{4} J-\frac{3\sqrt{3}}{10} \tilde{A}_1 +\frac{3}{100}( \tilde{C}_1+3\tilde{C}_2 +3\tilde{C}_3 +3\tilde{C}_4),\rule{5mm}{0mm}
\mathcal{B}_\parallel = \frac{1}{4} B_\parallel +\frac{9}{4} J +\frac{3\sqrt{3}}{10} \tilde{A}_1 +\frac{3}{50}(2\tilde{C}_1+3\tilde{C}_3),
\notag\\
&\mathcal{B}' = \frac{1}{4}B' +\frac{9}{10} \tilde{A}_2 -\frac{3\sqrt{3}}{50} \tilde{D}_1 +\frac{9}{100}\tilde{D}_2,\rule{5mm}{0mm}
\mathcal{B}'' = \frac{1}{4}B'' -\frac{9}{100}(\sqrt{3}\tilde{E}_1 +\tilde{E}_2 +\tilde{E}_3),\notag\\
&\mathcal{C}_1 = \frac{1}{100} C_1 +\frac{3}{50} \tilde{B}_\perp,\rule{5mm}{0mm}
\mathcal{C}_2 = \frac{1}{100} C_2 +\frac{9}{50} \tilde{B}_\perp,\rule{5mm}{0mm}
\mathcal{C}_3 = \frac{1}{100} C_3 +\frac{9}{100} \tilde{B}_\perp +\frac{9}{100} \tilde{B}_\parallel,\;\;
\mathcal{C}_4 = \frac{1}{100} C_4 +\frac{9}{50} \tilde{B}_\perp +\frac{3}{25}\tilde{B}_\parallel,\notag\\
&\mathcal{D}_1 = \frac{1}{100} D_1 -\frac{3\sqrt{3}}{50}\tilde{B}',\;\;
\mathcal{D}_2 = \frac{1}{100} D_2 +\frac{9}{100}\tilde{B}',\notag\\
& \mathcal{E}_1 = \frac{1}{100} E_1 -\frac{9\sqrt{3}}{100}\tilde{B}'',\rule{5mm}{0mm}
\mathcal{E}_2 = \frac{1}{100} E_2 +\frac{9}{100}\tilde{B}'',\rule{5mm}{0mm}
\mathcal{E}_3 = \frac{1}{100} E_3 -\frac{9}{100}\tilde{B}''. 
\end{align}
Compared to $\mathcal{B}$ which is the coefficients of two dipolar operators, 
the operators which include one quadrupole have coefficients $\mathcal{A}$ that are by one order of magnitude smaller. 
The operators consisting of two quadrupoles have coefficients, $\mathcal{C}$, $\mathcal{D}$, $\mathcal{E}$, 
which are by two orders of magnitude smaller. 
Therefore, we confine ourselves to the exchange interactions between dipoles with only $\mathcal{B}$ terms. 
\par
Finally, the $\mathcal{B}$ terms in Eq.(\ref{eq:soexchange2}) are converted from the global frame to the local frame 
$\{\hat{\vb*{x}}_\mu,\hat{\vb*{y}}_\mu,\hat{\vb*{z}}_\mu \}$, and we reach the effective spin model Eq.(\ref{eq:Ham}) 
in the main text. 
\end{widetext}
\section{Calculation of exchange parameters}\label{sec:calculation}
In this Appendix, we evaluate the values of exchange parameters in Eq.~(\ref{eq:Ham}) for materials, 
using the equations in Appendix \ref{sec:Kugel-Khomskii} and \ref{sec:SOC_trigonal}. 

\begin{figure*}[hbt]
	\begin{center}
		\includegraphics[width=17cm]{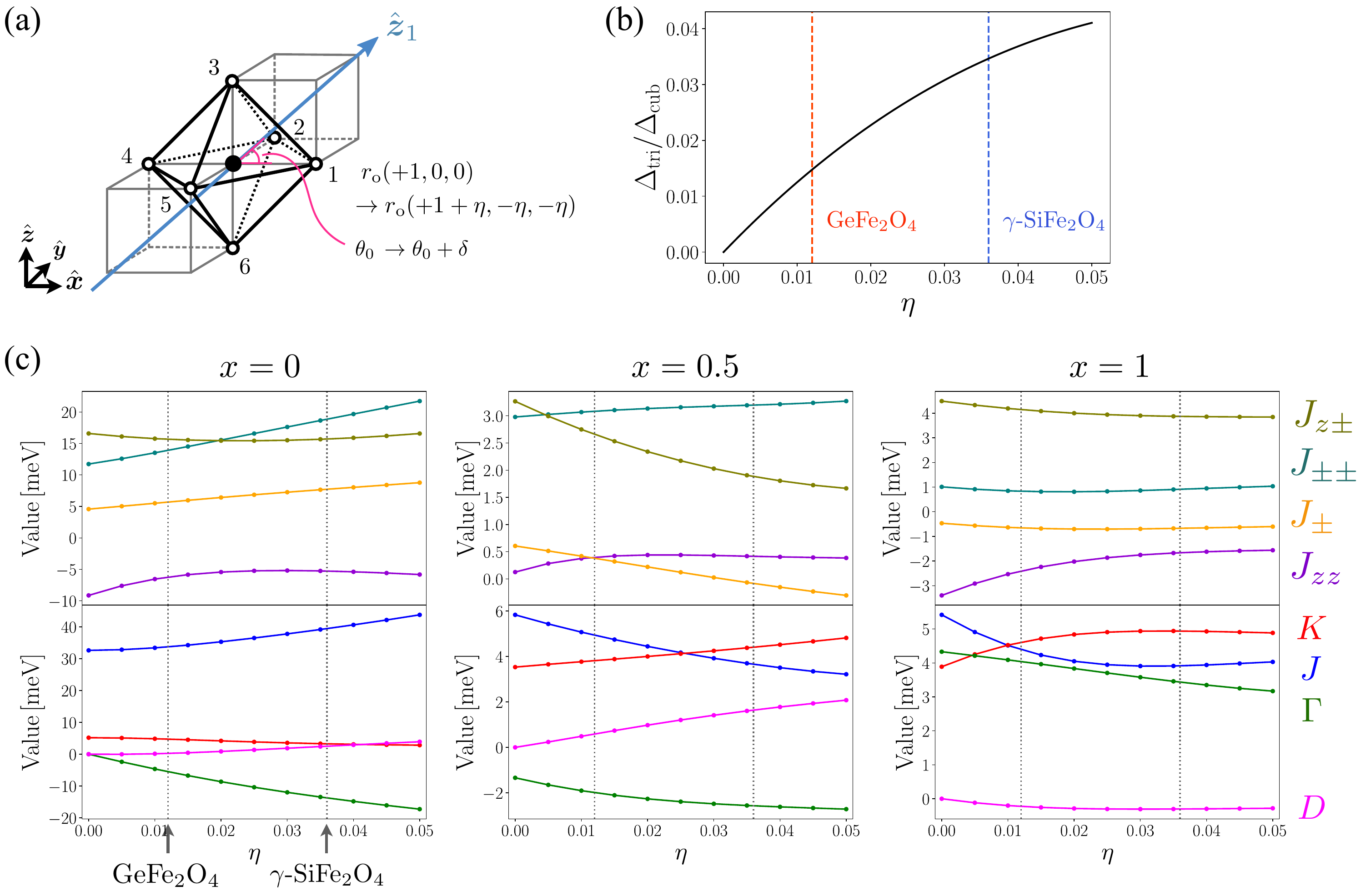}
		\caption{(a) Octahdron surrounding the magnetic ion (black circle) at sublattice $\mu=1$. The ligand ions (white circles) are compressed along the $\hat{\vb*{z}}_\mu$ axis. (b) The ratio of trigonal/cubic crystal field splitting $\Delta_{\rm tri}/\Delta_{\rm cub}$ as a function of the deviation parameter $\eta$. The red (blue) dashed line corresponds to $\eta$ for $\rm GeFe_2O_4$ ($\gamma$-$\rm SiFe_2O_4$). (c) The exchange parameters $\{J_{zz}, J_{\pm}, J_{\pm\pm}, J_{z\pm}\}$ in the local frame (top) and $\{J, K, \Gamma, D\}$ in the global frame for oxygen-mediated hopping only ($x=0$), 
			equivalently included oxygen-mediated/direct hopping ($x=0.5$), and direct hopping only ($x=1$). }
		\label{fig:S3}
	\end{center}
\end{figure*}
\subsection{Single-ion parameters}
We first outline the material parameters for on-site atomic Hamiltonian (\ref{eq:single-ion}). 
The Coulomb interaction parameters (\ref{eq:Coulomb_Coury}) are determined as follows: 
Part of the Slater-Condon parameters for the $\rm Fe^{2+}$ ion are known, 
$F^2=9.7\, \rm eV$ and $F^4=6.1\, \rm eV$ \cite{Tondello1967}, 
and the corresponding Coury parameters are, $J_{\rm C}=0.935\, \rm eV$ and $\Delta J_{\rm C}=0.129\, \rm eV$.
The remaining parameter $U_{\rm C}$ cannot be determined uniquely 
due to the uncertainty in the Slater-Condon parameter $F^0$.
Since $U_{\rm C}$ corresponds to the Kanamori parameter $U'_{\rm K}$, 
which represents the interorbital interaction, we expect it to be typically $\sim 3$ eV. 
\par
With the above information, we take $J_{\rm C}=1.0\, \rm eV$ and $\Delta J_{\rm C}=0.1\, \rm eV$. 
We set the cubic crystal field to $\Delta_{\rm cub} = 1.0 \, \rm eV$ and the SOC to $\lambda= 0.020 \, \rm eV$, 
which are roughly estimated values from $\Delta_{\rm cub} = 1.26 \, \rm eV$ \cite{Strobel1980} and $\lambda= 0.013 \, \rm eV$ \cite{Dunn1961,Walisinghe2021}, respectively. 

\subsection{Inter-atomic parameters}
For $\mathcal{H}_{\rm kin}$ in Eq.(\ref{eq:hkin}), 
there are two kinds of hoppings; the oxygen-mediated hopping and direct hopping between the magnetic ions. 
They are expressed by Slater-Koster parameters ($t_{pd\sigma}$, $t_{pd\pi}$, $t_{dd\sigma}$, $t_{dd\pi}$, $t_{dd\delta}$) \cite{Slater1954} 
and the charge transfer energy $\Delta_{pd}$ between $p$ orbitals of ligand ions and $d$ orbitals of magnetic ions. 
We set the hopping amplitudes as $t_o = t_{pd\pi}^2/\Delta_{pd} = 0.40 , \rm eV$ and $t_d = t_{dd\sigma} = -0.40 , \rm eV$ and assume $t_{pd\sigma}:t_{pd\pi}=-\sqrt{3}:1$ and $t_{dd\sigma}:t_{dd\pi}:t_{dd\delta}=6:-4:1$, following the typical treatment. 

To examine the effect of the two contributions, we rewrite the hopping Hamiltonian as
\begin{equation}
\mathcal{H}_{\rm kin}(x) = (1-x) \mathcal{H}^{\rm oxy}_{\rm kin} +x \mathcal{H}^{\rm dir}_{\rm kin}, 
\end{equation}
where $0 \leq x \leq 1$. 

\subsection{Trigonal distortion effect}\label{sec:calculation-3}
We now illustrate how the trigonal distortion of the ligand octahedron is treated and reflected 
in the parameters when taking a projection. 
The effect of this distortion is included in $\mathcal{H}_{\rm CEF}^{\rm tri}$ and $\mathcal{H}_{\rm kin}^{\rm oxy}$. 
When the magnetic ion is positioned at the origin, the positional vectors of the surrounding six ligand ions are given as 
\begin{equation}
\begin{split}
&\vb*{r}_{\rm o1} = r_{\rm o}(+1 +\eta, -\eta, -\eta), 
\quad \vb*{r}_{\rm o2} = r_{\rm o}(-\eta, +1+\eta -\eta), \\
&\vb*{r}_{\rm o3} = r_{\rm o}(-\eta, -\eta, +1+\eta), 
\quad \vb*{r}_{\rm o4} = r_{\rm o}(-1 -\eta, +\eta, +\eta), \\
&\vb*{r}_{\rm o5} = r_{\rm o}(+\eta, -1-\eta, +\eta), 
\quad \vb*{r}_{\rm o6} = r_{\rm o}(+\eta, +\eta, -1-\eta), 
\end{split}
\end{equation}
where $r_{\rm o}$ is the distance between the magnetic ion and ligand ions without distortion
and $\eta$ represents the displacement from the cubic structure [see Fig.~\ref{fig:S3}(a)]. 
When the octahedron is compressed/elongated in the direction of the $C_{3v}$-rotational axis, 
$\eta$ is positive/negative. 
We take $\eta=0.012$ for $\rm GeFe_2O_4$ and $\eta=0.036$ for $\gamma$-$\rm SiFe_2O_4$ \cite{Perversi2018}. 
The ratio of crystal field splitting $\Delta_{\rm tri}/\Delta_{\rm cub}$ is calculated 
using a point charge approximation to CEF \cite{Khomskii2016}. 
For $\rm GeFe_2O_4$, we find $\Delta_{\rm tri}/\Delta_{\rm cub}=0.015$, 
and for $\gamma$-$\rm SiFe_2O_4$, $\Delta_{\rm tri}/\Delta_{\rm cub}=0.035$ [see Fig.~\ref{fig:S3}(b)]. 
\par
The degree of trigonal distortion is parameterized by the angle $\theta_0+\delta$ between the  
the trigonal axis and the bond from-metal-to-ligand, 
where $\theta_0$ corresponds to the angle for the ideal octahedron. 
This angle is connected with the distortion parameter $\eta$ as
\begin{equation}
\cos(\theta_0 +\delta) = \frac{1-\eta}{\sqrt{3(1+2\eta+3\eta^2)}}. 
\end{equation}
The value for $\rm GeFe_2O_4$ corresponding to $\eta=0.012$ is $\delta=0.96^\circ$. 
This distortion is much smaller than pyrochlore oxide $\rm Eu_2Ir_2O_7$ with $\delta=5.9^\circ$\cite{Uematsu2015}, 
whose CEFs yield $\Delta_{\rm tri}/\Delta_{\rm cub} \simeq 0.13$. 
\par
With this in mind, we reexamine the previous literature \cite{Strobel1980} evaluating 
the values of CEF as $\Delta_{\rm tri} = 118 \, \rm meV$ from the effective Bohr magneton 
of the magnetic susceptibility measurement and find that this value is unlikely. 
Because, combined with $\Delta_{\rm cub} = 1.26 \, \rm eV$ from the optical absorbtion measurement\cite{Strobel1980}, 
it yields $\Delta_{\rm tri}/\Delta_{\rm cub} \simeq 0.09$, which is comparably as large as 
the largely distorted $\rm Eu_2Ir_2O_7$, and is unrealistic. 
\par 
Using the values, $\Delta_{\rm tri}/\Delta_{\rm cub}=0.015$ and $\Delta_{\rm cub} = 1.26 \, \rm eV$, 
we obtain $\Delta_{\rm tri} \sim 19 \, \rm meV$. 
\par
The effect of trigonal distortion on $\mathcal{H}_{\rm kin}^{\rm oxy}$ is included using Slater-Koster parameters. 
\smallskip\smallskip

\subsection{Results}
Let us present the model parameters obtained as functions of these material parameters. 
Figure \ref{fig:S3}(c) shows the exchange parameters for the three cases; 
oxygen-mediated hoppings only ($x=0$), oxygen-mediated/direct hoppings being equivalent ($x=0.5$), 
and direct hopping only ($x=1$). 
To realize the amplitude wave ordering, we need larger values of $J_{zz}$ and $J_{z\pm}$ compared to $J_{\pm}$ and $J_{\pm\pm}$. 
From our results, we find that both $J_{\pm}$ and $J_{\pm\pm}$ are more suppressed at $x=1$ than at $x=0$, 
indicating that the $x=1$ with only the direct hopping is more desirable. 
See also Table~\ref{table:local} in the main text. 
\par
There are two conventions to describe the effective spin Hamiltonian. 
One is to take the spin operators in the sublattice $\mu$ as 
$\mathsf{\vb*{S}}_\mu=(\mathsf{S}_\mu^x, \mathsf{S}_\mu^y, \mathsf{S}_\mu^z)$ in the local frame, 
and the other is to express it in the global frame where we introduce them as 
$\tilde{\vb*{S}}_\mu =(\tilde{S}_\mu^x, \tilde{S}_\mu^y, \tilde{S}_\mu^z)$. 
The local frame operators are connected to the one in the global frame as 
\begin{equation}
\tilde{\vb*{S}}_\mu =\mathsf{S}_\mu^x \hat{\vb*{x}}_\mu +\mathsf{S}_\mu^y \hat{\vb*{y}}_\mu +\mathsf{S}_\mu^z \hat{\vb*{z}}_\mu. 
\end{equation} 
and the form of effective Hamiltonian on the $z$ bond is expressed in the global frame as
\begin{equation}
\mathcal{H}_{14}^{z, \rm eff} = (\tilde{S}_1^x, \tilde{S}_1^y, \tilde{S}_1^z) \mqty(J & \Gamma & -D/\sqrt{2} \\ \Gamma & J & -D/\sqrt{2} \\ +D/\sqrt{2} & +D/\sqrt{2} & J+K) \mqty(\tilde{S}_4^x \\ \tilde{S}_4^y \\ \tilde{S}_4^z), 
\end{equation}
where $J$, $K$, $\Gamma$, and $D$ are the Heisenberg interaction, Kitaev interaction, symmetric off-diagonal exchange anisotropy, and the Dzyaloshinskii-Moriya interaction, respectively. 
These parameters are related to the local frame parameters as Eq.(\ref{eq:local-to-global}) in the main text, 
and their values in the present case are shown in Fig.~\ref{fig:S3}(c). 
\par
Notably, $J$, $K$, and $\Gamma$ are comparable in the preferable situation for the amplitude wave ordering, $x=1$, 
as shown in Fig.~\ref{fig:S3}(c). 
On the other hand, at $x=0$, it is found that $J \gg K,\Gamma,D$, indicating that it is relatively close to the Heisenberg model. 
This tendency is in sharp contrast to the heavy transition metal magnets with edge-shared octahedra, where the anisotropic exchange interactions are mostly mediated by the indirect hopping $(x=0)$, such as Kitaev honeycomb materials. 
Some representative values for the exchange parameters \cite{Songvilay2020,Winter2016} are listed in 
Table~\ref{table:local}. 

\section{Point group symmetry and order parameters}\label{sec:point_group}
\begin{table}[hbt]
	\caption{Character table of (a) $T_d$,  (b) $C_{3v}$, and (c) $C_{2v}$. }
	\label{table:character}
        \begin{minipage}{0.4\textwidth}
		\begin{ruledtabular}
			\begin{tabular}{rrrrrrr}
	                (a)\;\; $T_d$ & $\varepsilon$ & $8C_3$ & $3C_2$ & $6S_4$ & $6\sigma_d$ \\ \hline
				$A_1$ & $1$           & $1$    & $1$    & $1$    & $1$         \\
				$A_2$ & $1$           & $1$    & $1$    & $-1$   & $-1$        \\
				$E$   & $2$           & $-1$   & $2$    & $0$    & $0$         \\
				$T_1$ & $3$           & $0$    & $-1$   & $1$    & $-1$        \\
				$T_2$ & $3$           & $0$    & $-1$   & $-1$   & $1$       \\ 
			\end{tabular}
       \rule{0cm}{1mm}
			\begin{tabular}{rrrr}
			 (b)\;\; $C_{3v}$ & $\varepsilon$ & $2C_3$ & $3\sigma_v$ \\ \hline
				$A_1$    & $1$           & $1$    & $1$         \\
				$A_2$    & $1$           & $1$    & $-1$        \\
				$E$      & $2$           & $-1$   & $0$         \\ 
			\end{tabular}
      \rule{0cm}{1mm}
			\begin{tabular}{rrrrr}
			 (c)\;\; $C_{2v}$ & $\varepsilon$ & $C_2$ & $\sigma_{v}$ & $\sigma_{v'}$ \\ \hline
				$A_1$    & $1$           & $1$   & $1$           & $1$           \\
				$A_2$    & $1$           & $1$   & $-1$          & $-1$          \\
				$B_1$    & $1$           & $-1$  & $1$           & $-1$          \\
				$B_2$    & $1$           & $-1$  & $-1$          & $1$          \\ 
			\end{tabular}
		\end{ruledtabular}
	\end{minipage}
\end{table}
\begin{table*}[tbp]
	\caption{Spin order parameters of (a) $T_{d}$, (b)$C_{3v}$, and (c) $C_{2v}$. The spin operators are defined in the local frame.}
	\label{table:spin}
	\begin{minipage}{0.75\textwidth}
		\begin{ruledtabular}
(a) $T_{d}$ \rule{12cm}{0cm} \rule{0cm}{5mm}
			\begin{tabular}{lc}
				$A_2$    & $\frac{1}{2} (S_1^z +S_2^z +S_3^z +S_4^z)$              \\
				$E$    & $\frac{1}{2}\mqty(S_1^x +S_2^x +S_3^x +S_4^x 
				\\ S_1^y +S_2^y +S_3^y +S_4^y)$          \\
				$T_1^{(1)}$    & $\frac{1}{2}\mqty(S_1^z +S_2^z-S_3^z-S_4^z \\ 
				S_1^z -S_2^z+S_3^z-S_4^z \\
				S_1^z -S_2^z-S_3^z+S_4^z)$          \\
				$T_1^{(2)}$    & $\frac{1}{4}\mqty(2(S_1^x +S_2^x-S_3^x-S_4^x) \\ 
				-S_1^x+\sqrt{3}S_1^y+S_2^x-\sqrt{3}S_2^y-S_3^x+\sqrt{3}S_3^y+S_4^x-\sqrt{3}S_4^y \\
				-S_1^x-\sqrt{3}S_1^y+S_2^x+\sqrt{3}S_2^y+S_3^x+\sqrt{3}S_3^y-S_4^x-\sqrt{3}S_4^y)$          \\
				$T_2$    & $\frac{1}{4}\mqty(2(S_1^y +S_2^y-S_3^y-S_4^y) \\ 
				-\sqrt{3}S_1^x-S_1^y+\sqrt{3}S_2^x+S_2^y-\sqrt{3}S_3^x-S_3^y+\sqrt{3}S_4^x+S_4^y \\
				\sqrt{3}S_1^x-S_1^y-\sqrt{3}S_2^x+S_2^y-\sqrt{3}S_3^x+S_3^y+\sqrt{3}S_4^x-S_4^y)$          \\
			\end{tabular}
(b) $C_{3v}$ \rule{12cm}{0cm} \rule{0cm}{5mm}
			\begin{tabular}{lcc}
				& $\mu=1$ & $\mu=2,3,4$  \\ \hline
				$A_1$    & $0$           & $\frac{1}{2\sqrt{3}} (2S_2^y -\sqrt{3}S_3^x-S_{3}^y+\sqrt{3}S_4^x-S_4^y)$   \\
				$A_2$    & $S_1^z$         & $\frac{1}{\sqrt{3}} (S_2^z +S_3^z +S_4^z), \quad \frac{1}{2\sqrt{3}} (2S_2^x -S_3^x +\sqrt{3}S_3^y -S_4^x -\sqrt{3}S_4^y)$  \\
				$E$    & $\mqty( S_1^x \\ S_1^y)$           & 
				\begin{tabular}{c}
					$\frac{1}{6} \mqty(4S_2^x -2\sqrt{2}S_2^z +S_3^x -\sqrt{3}S_3^y +\sqrt{2}S_3^z +S_4^x +\sqrt{3}S_4^y +\sqrt{2}S_4^z \\
					-\sqrt{3}S_3^x+3S_3^y-\sqrt{6}S_3^z +\sqrt{3}S_4^x+3S_4^y+\sqrt{6}S_4^z)$  \\
					$\frac{1}{6}\mqty(-S_2^x-\sqrt{3}S_2^y-\sqrt{2}S_2^z -4S_3^x+2\sqrt{2}S_3^z -S_4^x+\sqrt{3}S_4^y-\sqrt{2}S_4^z \\
					-\sqrt{3}S_2^x-3S_2^y-\sqrt{6}S_2^z+\sqrt{3}S_4^x-3S_4^y+\sqrt{6}S_4^z)$ \\
					$\frac{1}{6}\mqty(-S_2^x+\sqrt{3}S_2^y-\sqrt{2}S_2^z-S_3^x-\sqrt{3}S_3^y-\sqrt{2}S_3^z-4S_4^x+2\sqrt{2}S_4^z \\
					\sqrt{3}S_2^x-3S_2^y+\sqrt{6}S_2^z-\sqrt{3}S_3^x-3S_3^y-\sqrt{6}S_3^z)$
				\end{tabular}
				\\
			\end{tabular}
(c) $C_{2v}$\rule{12cm}{0cm} \rule{0cm}{5mm}
			\begin{tabular}{lcc}
				& $\mu=1,2$ & $\mu=3,4$  \\ \hline
				$A_1$    & $\frac{1}{\sqrt{2}} (S_1^y +S_2^y)$           & $\frac{1}{\sqrt{2}} (S_3^y +S_4^y)$   \\
				$A_2$    & $\frac{1}{\sqrt{2}} (S_1^x +S_2^x),\quad \frac{1}{\sqrt{2}} (S_1^z +S_2^z)$         & $\frac{1}{\sqrt{2}} (S_3^x +S_4^x), \quad \frac{1}{\sqrt{2}} (S_3^z +S_4^z)$  \\
				$B_1$    & $\frac{1}{\sqrt{2}} (S_1^y -S_2^y)$           & $\frac{1}{\sqrt{2}} (S_3^x -S_4^x),\quad \frac{1}{\sqrt{2}} (S_3^z -S_4^z)$  \\
				$B_2$    & $\frac{1}{\sqrt{2}} (S_1^x -S_2^x), \quad \frac{1}{\sqrt{2}} (S_1^z -S_2^z)$           & $\frac{1}{\sqrt{2}} (S_3^y -S_4^y)$   \\ 
			\end{tabular}
		\end{ruledtabular}
	\end{minipage}
\end{table*}

In the mean-field approximation, the magnetic order parameters within a tetrahedron play a crucial role in determining the magnetic phases of pyrochlore magnets. In particular, to clearly characterize the amplitude wave phases, it is necessary to consider magnetic order parameters in cases where the sublattices become inequivalent. 
This section provides a list of those magnetic order parameters, which were omitted in the main text.

The symmetry operations for a single tetrahedron are as follows: 
(i) identity ($\varepsilon$); 
(ii) $2n\pi/3$ rotations around the axis $[111]$ and other six symmetry-related ones ($8C_{3}$); 
(iii) $\pi$ rotation around the axis $[100]$ and other two symmetry-related ones ($3C_{2}$); 
(iv) $(2n+1)\pi/2$ rotations around the axis $[100]$ and then the reflection in the plane perpendicular to the rotation axis and other four symmetry-related ones ($6S_{4}$); and
(v) reflection in the plane perpendicular to the axis $[110]$ and other five symmetry-related ones ($6\sigma_{d}$). 
The corresponding point group is $T_d$. 
Based on the character table of $T_d$ given in Table~\ref{table:character}(a), 
the spin operators on the tetrahedron are decomposed as $T_d = A_1 \oplus E \oplus 2T_1 \oplus T_2$. 
The explicit form of the order parameters expressed in the local frame is listed in Table~\ref{table:spin}(a). 

When the magnetic sites become inequivalent with the ratio of $1:3$, the point group $T_d$ is reduced to $C_{3v}$. 
Here we consider the case where the magnetic sites are divided into $\mu=1$ and $\mu=2,3,4$. 
The symmetry operations are (i) identity ($\varepsilon$); (ii) $2\pi/3$ rotations around the axis $[111]$ ($2C_{3}$); and (iii) reflection in the plane perpendicular to the axes $[1\bar{1}0]$, $[\bar{1}01]$, and $[01\bar{1}]$ ($3\sigma_{d}$). 
The corresponding point group is $C_{3v}$. 
Based on the character table of $C_{3v}$ given in Table~\ref{table:character}(b), the spin operators on the tetrahedron are decomposed as $C_{3v} = A_2 \oplus E $ for $\mu=1$ and $C_{3v} = A_1 \oplus 2A_2 \oplus 3E$ for $\mu=2,3,4$. 
The explicit form of the order parameters expressed in the local frame is listed in Table~\ref{table:spin}(b).  

When the magnetic sites become inequivalent with the ratio of $2:2$, the point group $T_d$ is reduced to $C_{2v}$. 
Here we consider the case where the magnetic sites are divided into $\mu=1,2$ and $\mu=3,4$. 
The symmetry operations are (i) identity ($\varepsilon$); (ii) $\pi$ rotation around the axis $[001]$ ($C_{2}$); (iii) reflection in the plane perpendicular to the axis $[1\bar{1}0]$ and $[110]$ ($\sigma_{v}$ and $\sigma_{v'}$, respectively). 
The corresponding point group is $C_{2v}$. 
Based on the character table of $C_{2v}$ given in Table~\ref{table:character}(c), the spin operators on the tetrahedron are decomposed as $C_{2v} = A_1 \oplus 2A_2 \oplus B_1 \oplus 2B_2$ for $\mu=1,2$ and $C_{3v} = A_1 \oplus 2A_2 \oplus 2B_1 \oplus B_2$ for $\mu=3,4$. 
The explicit form of the order parameters expressed in the local frame is listed in Table~\ref{table:spin}(c). 
\vspace{15mm}

\section{Description of a single spin-1 state}\label{sec:spin-1}
\subsection{SU(3) coherent state}
In this section, we consider a single $S=1$ state. 
The general spin-1 state is expressed by an SU(3) coherent state with four parameters, $\{\xi_1, \xi_2, \xi_3, \xi_4\}$: 
\begin{equation}\label{eq:su3}
\begin{split}
\ket*{\psi} &= e^{i\xi_3} \cos\xi_2\sin\xi_1 \ket*{+1} \\
&\quad +\cos\xi_1 \ket*{0} +e^{i\xi_4}\sin\xi_2\sin\xi_1 \ket*{-1}, 
\end{split}
\end{equation}
where $\ket*{\pm 1}$ and $\ket*{0}$ are eigenstates of $S^z$. 
The relationship between Eq.(\ref{eq:su3}) and the physical degrees of freedom is not obvious. 
To make it clear, we introduce the time-reversal-invariant basis, 
\begin{equation}
\begin{split}
&\ket*{x}=i\frac{\ket*{+1}-\ket*{-1}}{\sqrt{2}}, \\
&\ket*{y} = \frac{\ket*{+1}+\ket*{-1}}{\sqrt{2}}, \\
&\ket*{z}=-i\ket*{0}, 
\end{split}
\end{equation}
and represent a spin-1 state as 
\begin{equation}
\ket*{\psi} = \sum_{\alpha=x,y,z} (u_\alpha +i v_{\alpha}) \ket*{\alpha}, \quad u_{\alpha}, v_{\alpha} \in \mathbb{R}. 
\end{equation}
From $\{u_x, u_y, u_z, v_x, v_y, v_z\}$, four parameters are independent, 
as there are two constraints imposed; $u^2+v^2=1$ ($u=|\vb*{u}|$ and $v=|\vb*{v}|$) and $\vb*{u}\cdot\vb*{v}=0$. 
In this time-reversal-invariant basis, the spin operators and the quadrupole operators are represented as
\begin{eqnarray}
&& S^\alpha = -i\sum_{\beta\gamma} \varepsilon_{\alpha\beta\gamma} \ketbra*{\beta}{\gamma}, \\
&& Q^{\alpha\beta} = \frac{2}{3}\delta_{\alpha\beta} -\ketbra*{\beta}{\alpha}-\ketbra*{\alpha}{\beta}, 
\end{eqnarray}
respectively. 
The expectation values of these operators are given by
\begin{equation}
\ev*{\vb*{S}} = 2\vb*{u}\cross\vb*{v}, \quad \ev*{Q^{\alpha\beta}} = \frac{2}{3} \delta_{\alpha\beta} -2(u_\alpha u_\beta +v_\alpha v_\beta). 
\end{equation}
\smallskip
\smallskip
\subsection{Representation of SU(3) coherent state with uniaxial symmetry}
Now, since we are working on a case where the amplitude of the dipolar moment can vary between 0 and 1, 
it is convenient to parametrize the expectation value of the spin operators as 
\begin{equation} \label{eq:spin_mst}
\ev*{\vb*{S}} = M\mqty( \sin\theta\cos\varphi \\  \sin\theta\sin\varphi \\ \cos\theta). 
\end{equation}
From $u^2+v^2=1$ and $4u^2v^2=M^2$, $u$ and $v$ are expressed as 
\begin{equation}
u=\sqrt{\frac{1+\sqrt{1-M^2}}{2}}, \quad v=\sqrt{\frac{1-\sqrt{1-M^2}}{2}}, 
\end{equation}
where $u>v$ is assumed. 
To reconcile with this newly introduced description instead of using $\vb*{u}$ and $\vb*{v}$ vectors, 
we need to add one more parameter and make them four; $M$, $\theta$, $\varphi$, and $\eta$. 
To characterize them, we first consider the state where the dipole(spin) points in the $z$ direction, 
whose $\vb*{u}$ and $\vb*{v}$ vectors are given by
\begin{equation}
\vb*{u} = u\mqty( \cos\eta \\ \sin\eta \\ 0 ), \quad \vb*{v} = v\mqty( -\sin\eta \\ \cos\eta \\ 0). 
\end{equation}
Starting from this state, 
we make a $\theta$ rotation of $x$, $y$, $z$ axes about the vector $\hat{\vb*{n}}$ 
to obtain a dipole moment pointing in the direction $(\theta, \varphi)$ as 
\begin{equation}
\hat{\vb*{n}}=\frac{\hat{\vb*{z}}\cross\ev*{\vb*{S}}}{|\hat{\vb*{z}}\cross\ev*{\vb*{S}}|} = \mqty(-\sin\varphi \\ \cos\varphi \\ 0 ). 
\end{equation}
The corresponding $\vb*{u}$ and $\vb*{v}$ vectors are given by
\begin{equation}
\begin{split}
& \vb*{u} = u\mqty( \cos\eta\cos\theta -\sin\varphi\sin(\eta-\varphi)(1-\cos\theta) \\
\sin\eta\cos\theta +\cos\varphi\sin(\eta-\varphi)(1-\cos\theta) \\
-\cos(\eta-\varphi)\sin\theta), \\
& \vb*{v} = v\mqty( -\sin\eta\cos\theta-\sin\varphi\cos(\eta-\varphi)(1-\cos\theta) \\
\cos\eta\cos\theta+\cos\varphi\cos(\eta-\varphi)(1-\cos\theta) \\
\sin(\eta-\varphi)\sin\theta). 
\end{split}
\end{equation}
Using this expression, the expectation value of $S_z^2$ is given by
\begin{equation}
\begin{split}
\ev*{(S^z)^2} &= u_x^2+u_y^2 +v_x^2+v_y^2 = u^2-u_z^2 +v^2-v_z^2 \\
&= u^2 -u^2\cos^2(\eta-\varphi)\sin^2\theta \\
&\quad + v^2 -v^2\sin^2(\eta-\varphi)\sin^2\theta \\
&= 1- \sin^2\theta [u^2\cos^2(\eta-\varphi) +v^2\sin^2(\eta-\varphi)] \\
&= 1- \frac{1}{2}\sin^2\theta \Big[(1+\sqrt{1-M^2})\cos^2(\eta-\varphi)\\
 &\quad  +(1-\sqrt{1-M^2})\sin^2(\eta-\varphi) \Big] \\
&= 1- \frac{1}{2}\sin^2\theta \Big[1+\sqrt{1-M^2}\cos2(\eta-\varphi)\Big]. 
\end{split}
\end{equation}
In our spin Hamiltonian with easy-plane SIA, $D>0$, the quadrupole moment has a U(1) symmetry about the $z$ axis, in which case the parameter $\eta$ is uniquely determined as $\eta=\varphi$. 
In this case, the quadrupole moments are expressed by only $M$, $\theta$, $\varphi$ as
\begin{equation}
\begin{split}
& \ev*{Q^{3z^2-r^2}} = \frac{1}{2\sqrt{3}} \Big[ 2-3(1+\sqrt{1-M^2}) \sin^2\theta \Big], \\
& \ev*{Q^{x^2-y^2}} = \frac{1}{2} \Big(\sin^2\theta -(1+\cos^2\theta)\sqrt{1-M^2} \Big) \cos2\varphi, \\
& \ev*{Q^{xy}} = \frac{1}{2} \Big(\sin^2\theta -(1+\cos^2\theta)\sqrt{1-M^2} \Big) \sin2\varphi, \\
& \ev*{Q^{yz}} = \frac{1}{2} (1+\sqrt{1-M^2})\sin2\theta\sin\varphi, \\
& \ev*{Q^{zx}} = \frac{1}{2} (1+\sqrt{1-M^2})\sin2\theta\cos\varphi.
\end{split}
\end{equation}
\nocite{apsrev41Control}
\bibliographystyle{apsrev4-2}
\bibliography{brunogeierite}
	
\end{document}